\documentclass[11pt,a4paper]{article}
\usepackage{jheppub}
\usepackage{graphicx}
\usepackage{placeins}
\usepackage{dcolumn}
\usepackage{bm}
\usepackage{xspace}
\usepackage{subfig}
\usepackage{relsize}
\usepackage{multirow}
\usepackage{braket}
\usepackage{relsize}
\usepackage[mathlines]{lineno}
\usepackage{etoolbox}
\usepackage{ragged2e}
\usepackage{amsmath,slashed,bbold}
\usepackage{rotating}
\apptocmd{\thebibliography}{\justifying}{}{} 
\usepackage{lieart}
\usepackage{booktabs}
\usepackage{siunitx}
\usepackage{slashed}
\usepackage{amssymb}
\usepackage[usenames,dvipsnames,svgnames,table]{xcolor}
\DeclareMathAlphabet{\mathpzc}{OT1}{pzc}{m}{it}

\newcommand{\GeV}{\si{\GeV}}
\newcommand{\TeV}{\si{\TeV}}

\newcommand{\HEPfit}{\texttt{HEPfit}\xspace}

\usepackage{color}
\usepackage{hyperref}
\usepackage[all]{hypcap}
\hypersetup{
    unicode=false,          
    pdftoolbar=true,        
    pdfmenubar=true,        
    pdffitwindow=false,     
    pdfstartview={FitH},    
    pdftitle={B anomalies under the lens of electroweak precision},    
    pdfauthor={Lina Alasfar, Aleksandr Azatov, Jorge de Blas, Ayan Paul, Mauro Valli},     
    pdfsubject={hep-ph},   
    pdfkeywords={B Physics} {EWPO} {SMEFT}, 
    pdfnewwindow=true,      
    colorlinks=true,       
    linkcolor=blue,          
    citecolor=blue,        
    filecolor=magenta,      
    urlcolor=cyan,           
    linktocpage=true
}

\def\equationautorefname~#1\null{equation\,(#1)\null}
\newcommand{\appendixref}[1]{\hyperref[#1]{appendix~\ref{#1}}}

\begin{document}

\preprint{
\begin{flushright}
DESY 20-091\\
HU-EP-20/12-RTG\\
SISSA 16/2020/FISI \\
UCI-TR 2020-10
\end{flushright}
}

\title{\boldmath$B$ anomalies under the lens of electroweak precision}

\author[a,]{Lina Alasfar,}
\author[b,c]{Aleksandr Azatov,}
\author[d]{Jorge de Blas,}
\author[a,e]{Ayan Paul,}
\author[f]{and Mauro Valli}

\affiliation[a]{Institut f\"ur Physik, Humboldt-Universit\"at zu Berlin, D-12489 Berlin, Germany}
\affiliation[b]{SISSA International School for Advanced Studies, Via Bonomea 265, 34136, Trieste, Italy}
\affiliation[c]{INFN - Sezione di Trieste, Via Bonomea 265, 34136, Trieste, Italy}
\affiliation[d]{Institute of Particle Physics Phenomenology, Durham University, Durham DH1 3LE, UK}
\affiliation[e]{DESY, Notkestrasse 85, D-22607 Hamburg, Germany}
\affiliation[f]{Department of Physics and Astronomy, University of California, Irvine, CA 92697-4575 USA}

\emailAdd{alasfarl@physik.hu-berlin.de}
\emailAdd{aleksandr.azatov@sissa.it}
\emailAdd{jorge.de-blas-mateo@durham.ac.uk}
\emailAdd{ayan.paul@desy.de}
\emailAdd{mvalli@uci.edu}

\abstract{The measurements carried out at LEP and SLC projected us into the precision era of electroweak physics. This has also been relevant in the theoretical interpretation of LHCb and Belle measurements of rare $B$ semileptonic decays, paving the road for new physics with the inference of lepton universality violation in $R_{K^{(*)}}$ ratios. The simplest explanation of these flavour anomalies -- sizeable one-loop contributions respecting Minimal Flavour Violation -- is currently disfavoured by electroweak precision data. In this work, we discuss how to completely relieve the present tension between electroweak constraints and one-loop minimal flavour violating solutions to $R_{K^{(*)}}$. We determine the correlations in the Standard Model Effective Field Theory that highlight the existence of such a possibility. Then, we consider minimal extensions of the Standard Model where our effective-field-theory picture can be realized. We discuss how these solutions to $b \to s \ell \ell$ anomalies, respecting electroweak precision and without any new source of flavour violation, may point to the existence of a $Z^{\prime}$ boson at around the TeV scale, within the discovery potential of LHC, or to leptoquark scenarios.}

\maketitle
\section{Introduction}
\label{sec:intro}

In the era of the Large Hadron Collider (LHC) an intense program aimed at probing the Standard Model (SM) at the TeV scale {has been established}. At the same time, one of the most valuable sources for the study of new physics (NP) above the electroweak (EW) scale is provided by indirect tests of the SM via the so-called the EW precision observables (EWPO). These include, in particular, the very precise measurements at the $Z$ pole performed at the Large Electron-Positron (LEP) collider and the Stanford Linear Collider (SLC). In corroboration with the Higgs-boson discovery and the experimental information collected at LHC and Tevatron, they provide strong constraints on theories beyond the SM (BSM) that lead to important deformations of the standard EW sector~\cite{Falkowski:2013dza,Ciuchini:2013pca,Falkowski:2014tna,deBlas:2015aea,deBlas:2016ojx,deBlas:2017wmn,Haller:2018nnx,Ellis:2018gqa,Erler:2019hds,Dawson:2020oco}. Intriguingly, the interplay between the TeV region under scrutiny at the LHC and the NP probes represented by EW precision tests may be of fundamental importance for the study of the \textit{$B$-physics anomalies}~\cite{Bhattacharya:2014wla,Feruglio:2016gvd,Celis:2017doq,Buttazzo:2017ixm,Kumar:2018kmr,Ciuchini:2019usw,Aebischer:2019mlg,Cornella:2019hct}.

The outcome of LHCb and Belle analyses in the study of semileptonic $B$ decays points to the possible presence of NP in the measured ratios $R_{K^{(*)}} \equiv Br(B \to K^{(*)} \mu^{+} \mu^{-}) / Br(B \to K^{(*)} e^{+} e^{-})$ at low dilepton mass~\cite{Aaij:2014ora,Aaij:2017vbb,Aaij:2019wad,Abdesselam:2019wac}. The averaged experimental values deviate from unity at the $\sim2.5\sigma$ level, hinting at lepton universality violation (LUV). A statistically significant inference of LUV in $b \to s \ell \ell$ ($\ell = e, \mu$) transitions can be translated into a strong case for the evidence of BSM physics~\cite{Hiller:2014yaa,Hiller:2014ula,Bordone:2016gaq}.

The interpretation of these experimental results as an imprint of heavy new dynamics has primarily been assessed in a model-independent fashion via the language of effective field theories (EFT) in~\cite{DAmico:2017mtc,Geng:2017svp,Capdevila:2017bsm,Ciuchini:2017mik,Hiller:2017bzc} and more recently revisited in refs.~\cite{Ciuchini:2019usw,Aebischer:2019mlg,Alok:2019ufo,Alguero:2019ptt,Kowalska:2019ley,Arbey:2019duh,Datta:2019zca}. Furthermore, the NP picture depicted by these global analyses could also accommodate a set of tensions related to the well-measured muonic channel of these $B$ decays, in particular, to the angular analysis of $B \to K^{*} \mu^{+} \mu^{-}$~\cite{Descotes-Genon:2013wba,Descotes-Genon:2015uva}. These measurements have very recently been updated by the LHCb collaboration~\cite{Aaij:2020nrf}.

The set of tensions not related to LUV tests would specifically connect NP effects to muon-flavoured couplings. However, long-distant effects present in the amplitude of these processes~\cite{Khodjamirian:2010vf,Lyon:2014hpa,Chobanova:2017ghn,Blake:2017fyh,Bobeth:2017vxj} -- involving hadronic contributions that are theoretically difficult to handle~\cite{Jager:2014rwa,Ciuchini:2015qxb,Arbey:2018ics,Chrzaszcz:2018yza} -- make such a conclusion debatable, see, e.g.~\cite{Ciuchini:2018anp,Hurth:2020rzx}. From this point of view, the LUV information extracted from ratios of branching ratios and from observables like the ones considered in~\cite{Capdevila:2016ivx,Serra:2016ivr,Wehle:2016yoi,Alguero:2019pjc} remain the most promising avenue in the future for a more precise assessment of the overall tension seen in $b \to s \ell \ell$ measurements~\cite{Kou:2018nap}. Eventually, while a tighter upper limit has been recently obtained by LHCb on the branching ratio of $ B_{s} \to e^{+} e^{-}$~\cite{Aaij:2020nol}, the combined experimental average for the $ Br(B_{s} \to \mu^{+} \mu^{-}) $~\cite{Chatrchyan:2013bka,Aaij:2017vad,Aaboud:2018mst} also shows some tension with the SM prediction~\cite{Bobeth:2013uxa} as can be seen from the findings in~\cite{Ciuchini:2019usw,Aebischer:2019mlg}.

A broader discussion on $B$-physics anomalies should also include the LUV information stemming from another class of rare $B$ decays, namely $b \to c$ semileptonic transitions~\cite{Azatov:2018knx,Alok:2019uqc,Murgui:2019czp,Shi:2019gxi}. Indeed, a combined resolution of $R_{K^{(*)}}$ anomalies with the long-standing deviations observed in $R_{D^{(*)}} \equiv Br(B \to D^{(*)} \tau \nu) / Br(B \to D^{(*)} \ell \nu)$ originally found at Babar~\cite{Lees:2013uzd} and subsequently measured at Belle~\cite{Huschle:2015rga} and LHCb~\cite{Aaij:2017uff}, has triggered a lot of interest in the theory community. In particular, in order for NP effects to simultaneously account for a $\sim 20\%$ deviation in tree-level charged-weak decays and in loop-level flavour-changing neutral currents (FCNC), models with a highly non-trivial flavour structure are required~\cite{DiLuzio:2017vat,Calibbi:2017qbu,Bordone:2017bld,Barbieri:2017tuq,Assad:2017iib,Heeck:2018ntp,Fornal:2018dqn,Crivellin:2018yvo,Crivellin:2019dwb,Bordone:2019uzc}, often being at the edge of flavour physics constraints~\cite{Bona:2007vi,Silvestrini:2018dos} and collider bounds~\cite{Greljo:2017vvb,Baker:2019sli}.
So far, model building has been mainly put forward in the direction of UV-completing low-energy leptoquark benchmarks identified, for instance,  in refs.~\cite{Calibbi:2015kma,Dorsner:2016wpm,Buttazzo:2017ixm,Kumar:2018kmr,Cornella:2019hct}. 

It is important to acknowledge that the most up-to-date measurements of $R_{D^{(*)}}$ from the Belle collaboration -- obtained by fully reconstructing the $\tau$ particle via the hadronic~\cite{Hirose:2016wfn} and, more notably, leptonic~\cite{Abdesselam:2019dgh} decay modes -- turns out to be in good agreement with the SM~\cite{Bigi:2016mdz,Bernlochner:2017jka,Bigi:2017jbd,Jaiswal:2017rve}. This fact may cast some doubt on the effective role one should really attribute to $ b \to c $ transitions in the interpretation of the depicted \textit{$B$-physics crisis}.

Therefore, in light of the recent results from Belle and LHCb, it is timely for us to focus again on the $b \to s \ell \ell $ conundrum and reassess the solutions to $B$-physics anomalies that can be realized at one loop without any new source of flavour violation. The simplest resolution of these anomalies has been proposed in ref.~\cite{Kamenik:2017tnu}, extending the SM with a single new Abelian gauge group, together with the presence of top- and muon-partners, resulting in a top-philic $Z^\prime$ boson capable of evading present collider constraints~\cite{Fox:2018ldq} and responsible for the required LUV signatures.

Such a minimal model actually falls into a larger category pointed out in ref.~\cite{Celis:2017doq} through the language of the Standard Model Effective Field Theory (SMEFT), and subsequently elaborated upon in greater detail in the phenomenological study of ref.~\cite{Camargo-Molina:2018cwu}. 
 
 At the basis of this class of proposals, the notable attempt is twofold: 
 \begin{itemize}
     \item[\textit{i)}] Addressing the deviations in these FCNC processes with NP effects entering at one-loop level, as for SM amplitudes. This reduces the original multi-TeV domain of NP for $B$ anomalies~\cite{DiLuzio:2017chi} to energies closer to present and future collider reach.
     \item[\textit{ii)}] Avoiding the introduction of new sources of flavour violation beyond the SM Yukawa couplings, relaxing in this way, any restrictive flavour probe of NP {in a fashion similar to what is predicted in Minimal Flavour Violation (MFV)~\cite{Buras:2000dm,DAmbrosio:2002vsn,Kagan:2009bn}.}
  \end{itemize}
 
The aforementioned proposal shows a strong tension with $Z$-pole precision observables~\cite{Camargo-Molina:2018cwu,Efrati:2015eaa}. In ref.~\cite{Ciuchini:2019usw} it has been shown that even in the presence of large hadronic effects in the amplitude of $B \to K^{*} \mu^{+} \mu^{-}$, a tension of at the 3$\sigma$ level at least would persist between $B$ data and EWPO for muonic LUV effects, and an even stronger tension would be found in the case of LUV scenarios involving electron couplings.

This fact has been brought to light recently~\cite{Coy:2019rfr} to abandon \textit{ii)}, and reformulate the original proposal addressing $B$ anomalies at one loop adding specific {BSM sources of flavour violation in order to reconcile $B$ data with EW precision tests in this context.} However, as briefly advertised in ref.~\cite{Ciuchini:2019usw}, an important caveat of this EW tension versus $B$ anomalies concerns the assumption of no tree-level NP contributions to EWPO. 

In this work, we attempt, for the first time, to provide a broad exploration of the possible cross-talk of NP in the EW sector and in the flavour playground for  $b \to s \ell \ell$ transitions. Firstly, we revisit the standard EW analysis in the presence of leading-log one-loop contributions from the renormalization group equations (RGE) evolution of the operators in the SMEFT~\cite{Jenkins:2013zja,Jenkins:2013wua}. Then, we perform a joint fit to the comprehensive experimental set that includes EWPO in conjugation with the state-of-the-art measurements of semileptonic $B$ decays. Our EFT analysis targets heavy new dynamics that contributes to $b \to s \ell \ell$ at the loop level only through SMEFT RGE, involving the SM Yukawa couplings as the only sources of flavour violation in the resolution of $B$ anomalies.

Within our study, we systematically review novel correlations among gauge-invariant dimension-six operators that help us shed new light on the one-loop solutions to $B$ anomalies. Continuing in the spirit of the previous work done by some of us~\cite{Ciuchini:2015qxb,Ciuchini:2016weo,Ciuchini:2017mik,Ciuchini:2017gva,Ciuchini:2018xll,Ciuchini:2018anp,Ciuchini:2019usw}, we shall furnish our results in both a conservative and optimistic approach to the non-perturbative hadronic contributions which can significantly affect the conclusions on the NP effects at hand.

On the basis of the SMEFT picture obtained from our combined inspection of EW and flavour data, we proceed to refine simple UV models already considered in the literature~\cite{Kamenik:2017tnu,Fox:2018ldq,Celis:2017doq}. We corner the interesting parameter space of this refined class of models where EWPO are respected while $B$ anomalies can be addressed {at one loop without introducing new sources of flavour violation}. Eventually, we go on to discuss the complementary probes offered by collider searches.

The paper is organized as follows: in \autoref{sec:theory} we review the ingredients of our EFT analysis; in \autoref{sec:strategy} we detail the strategy adopted for our combined EW+flavour fit in the SMEFT, the results from which are collected in \autoref{sec:EFT_results}; in \autoref{sec:UVtoymodels} we discuss the most economic viable $Z^{\prime}$ model in relation to our EFT results and also mention possible alternative leptoquark scenarios. Our conclusions are summarized in \autoref{sec:sum}.

\section{Theoretical preamble}
\label{sec:theory}

Previous global analyses of $b \to s \ell \ell$ anomalies have highlighted the appearance of new dynamics at a scale of $\mathcal{O}(10)$~TeV for $\mathcal{O}(1)$ effective couplings encoding NP effects at the tree level~\cite{DAmico:2017mtc,Geng:2017svp,Capdevila:2017bsm,Ciuchini:2017mik,Hiller:2017bzc}. The mass gap with the weak scale, characterized by the Higgs vacuum expectation value (VEV) $v \approx 246$ GeV, justifies the BSM translation of these results in the gauge-invariant formalism of the SMEFT~\cite{Buchmuller:1985jz,Grzadkowski:2010es}. At dimension six, in an operator product expansion in inverse powers of the NP scale $\Lambda$, and working in the Warsaw basis~\cite{Grzadkowski:2010es}, the operators of interest for the explanation of these $B$ anomalies are~\cite{Celis:2017doq,Ciuchini:2019usw,Aebischer:2019mlg}: 
\begin{eqnarray}
\label{eq:tree_LUV_SMEFT}
O^{LQ^{(1)}}_{\ell \ell 23} & = & \bar{L}_{\ell} \gamma_{\mu}  L_{\ell} \, \bar{Q}_{2} \gamma^{\mu}  Q_{3} \ , \nonumber \\
O^{LQ^{(3)}}_{\ell \ell 23} & = & \bar{L}_{\ell} \gamma_{\mu} \tau^{A} L_{\ell} \, \bar{Q}_{2} \gamma^{\mu} \tau^{A} Q_{3} \ , \nonumber \\
O^{Qe}_{23 \ell \ell} & = & \bar{Q}_{2} \gamma_{\mu} Q_{3} \, \bar{e}_{\ell} \gamma^{\mu}  e_{\ell} \ ,\nonumber \\
O^{Ld}_{ \ell \ell 23} & = & \bar{L}_{\ell} \gamma_{\mu} L \, \bar{d}_{2} \gamma^{\mu} d_{3} \ , \\
O^{ed}_{\ell \ell 23 } & = & \bar{d}_{2} \gamma_{\mu} d_{3} \, \bar{e}_{\ell} \gamma^{\mu}  e_{\ell} \ ,\nonumber
\end{eqnarray}
where weak doublets are represented in upper case, $SU(2)_{\rm L}$ singlets in lower case, and Pauli matrices $\tau^{A}$ characterize $SU(2)_{\rm L}$ triplet currents. Within available light-cone sum-rule results on long-distance effects in $B \to K^{*} \mu^{+} \mu^{-}$~\cite{Khodjamirian:2010vf,Bobeth:2017vxj}, data point to the presence of both the operators with $b \to s$ left-handed and right-handed currents with muonic flavour ($\ell = 2$) in eq.~\eqref{eq:tree_LUV_SMEFT}~\cite{Ciuchini:2019usw,Alok:2019ufo,Alguero:2019ptt,Kowalska:2019ley}. However, it is important to observe that: 
\begin{itemize}
\item The current statistical significance for the need of right-handed $b \to s$ couplings remain small, hinted only by the ratio $R_{K^{*}}/R_{K} \neq 1$ at the $1\sigma$ level~\cite{Hiller:2017bzc,Ciuchini:2019usw}. Hence, the present $B$ anomalies can be essentially addressed by $O^{LQ^{(1,3)}}_{22 23}$ and $O^{Qe}_{23 22}$.
\item Within a conservative approach to hadronic uncertainties~\cite{Jager:2014rwa,Ciuchini:2015qxb,Arbey:2018ics}, the preference for muonic NP effects in global analyses gets mitigated to a large extent and electro-phillic scenarios become viable too~\cite{Ciuchini:2017mik}; moreover, the fully left-handed operator(s)\footnote{The most promising observables that will allow  to genuinely disentangle NP effects in the future in the fully left-handed operator $O^{LQ^{(3)}}_{\ell \ell 23}$ from the ones of $O^{LQ^{(1)}}_{\ell \ell 23}$, are $B \to K^{(*)} \nu \bar{\nu}$ decays~\cite{Altmannshofer:2009ma,Buras:2014fpa,Descotes-Genon:2020buf}.} $O^{LQ^{(1,3)}}_{\ell \ell 23}$ offers the minimal model-independent resolution to $b \to s$ anomalies~\cite{Ciuchini:2019usw}.
\end{itemize}

Interestingly, with a leading expansion in the top-quark Yukawa coupling of the RGE computed in~\cite{Jenkins:2013zja,Jenkins:2013wua}, the Wilson coefficients associated to $O^{LQ}_{22 23}$ and $O^{Qe}_{23 22}$ can be generated at one loop by two distinct sets of dimension-six operators~\cite{Celis:2017doq} that can lead to LUV effects in $b \to s \ell \ell$ amplitudes without flavour violation in the quark current. A first set involves operators built of Higgs and leptonic currents:
\begin{eqnarray} 
\label{eq:SMEFT_op_HL}
O^{HL^{(1)}}_{\ell \ell} &=& ( H^{\dagger} i \overset{\leftrightarrow}{D}_{\mu}H ) (\bar{L}_{\ell} \gamma^{\mu}  L_{\ell} )\,, \nonumber \\
O^{HL^{(3)}}_{\ell \ell} &=& ( H^{\dagger} i \overset{\, \leftrightarrow_A}{D_{\mu}}H ) (\bar{L}_{\ell} \gamma^{\mu} \tau^{A} L_{\ell} )\,, \nonumber \\
O^{He}_{\ell \ell} &=& ( H^{\dagger} i \overset{\leftrightarrow}{D}_{\mu}H ) (\bar{e}_{\ell} \gamma^{\mu}  e_{\ell} )\,.
\end{eqnarray}
A second one corresponds to semileptonic four-fermion (SL-4F) operators with right-handed top-quark currents:
\begin{eqnarray} 
\label{eq:SMEFT_op_loop_lu}
O^{Lu}_{\ell \ell 3 3} &=& (\bar{L}_{\ell} \gamma_\mu L_{\ell})(\bar{u}_{3}\gamma^\mu u_{3})\,, \nonumber \\
O^{eu}_{\ell \ell 3 3} &=& (\bar{e}_{\ell} \gamma_\mu e_{\ell})(\bar{u}_{3}\gamma^\mu u_{3})\,.
\end{eqnarray}

Solving the RGE in  a leading-logarithmic approximation, the matching conditions for the left-handed quark-current operators in eq.~\eqref{eq:tree_LUV_SMEFT} at the scale $\mu_{\textrm{\tiny{EW}}}\sim v $ are:\footnote{In this work, for one-loop effects,  we assume the NP scale to be $\Lambda = 1$~TeV. We also set $\mu_{\rm EW} = m_t\simeq v/\sqrt{2}$ to minimize the matching-scale dependence with the inclusion of next-to-leading corrections~\cite{Aebischer:2015fzz,Bobeth:2017xry}.}
\begin{eqnarray}
\label{eq:SMEFT_matching_1loop}
C^{LQ^{(1)}}_{\ell \ell 23} &=& V_{ts}^{*} V^{ }_{tb} \left(\frac{y_{t} }{4 \pi}\right)^2 \log \left( \frac{\Lambda}{\mu_{\textrm{\tiny{EW}}}} \right)   \, \left(C^{Lu}_{\ell \ell 3 3} - C^{HL^{(1)}}_{\ell \ell} \right)\,,\nonumber \\
C^{LQ^{(3)}}_{\ell \ell 23} &=& V_{ts}^{*} V^{ }_{tb} \left(\frac{y_{t} }{4 \pi}\right)^2 \log \left( \frac{\Lambda}{\mu_{\textrm{\tiny{EW}}}} \right)   \,   C^{HL^{(3)}}_{\ell \ell} \,,\nonumber \\
C^{Qe}_{23\ell \ell} &=& V_{ts}^{*} V^{ }_{tb} \left(\frac{y_{t} }{4 \pi}\right)^2 \log \left( \frac{\Lambda}{\mu_{\textrm{\tiny{EW}}}} \right)   \, \left( C^{eu}_{\ell \ell 3 3} - C_{\ell \ell}^{He}\right) \, .
\end{eqnarray}

In terms of vectorial and axial currents typically discussed in the context of the weak effective theory at low energies~\cite{Buchalla:1995vs,Buras:1998raa,Silvestrini:2019sey}, the operators in eq.~\eqref{eq:SMEFT_matching_1loop} are matched to
\begin{eqnarray}
\label{eq:_Q9_Q10}
    O_{9 V, \ell} & = & \frac{\alpha_{e}}{8 \pi} (\bar{s} \gamma_{\mu} (1-\gamma_{5})b) ( \bar{\ell} \gamma^{\mu} \ell ) \nonumber \ , \ \\
    O_{10 A, \ell} & = & \frac{\alpha_{e}}{8 \pi} (\bar{s} \gamma_{\mu} (1-\gamma_{5})b) ( \bar{\ell} \gamma^{\mu} \gamma_{5} \ell ) \ ,
\end{eqnarray}
so that the matching conditions at the scale $\mu_{\textrm{\tiny{EW}}}$ for the set of operators in eq.~\eqref{eq:SMEFT_op_HL}~-~\eqref{eq:SMEFT_op_loop_lu} follow:
\begin{eqnarray} 
\label{eq:_C9_C10}
C_{9,\ell}^{\rm NP}&=& \frac{\pi v^{2}}{\alpha_{e} \Lambda^2} \left(\frac{y_{t} }{4 \pi}\right)^2 \log \left( \frac{\Lambda}{\mu_{\textrm{\tiny{EW}}}} \right)   \, \left(C^{HL^{(3)}}_{\ell \ell} - C^{HL^{(1)}}_{\ell \ell} -C^{He}_{\ell \ell} + C^{Lu}_{\ell \ell 3 3} + C^{eu}_{\ell \ell 3 3} \right)\,,\nonumber \\
C_{10,\ell}^{\rm NP}&=& \frac{\pi v^{2}}{\alpha_{e} \Lambda^2} \left(\frac{y_{t} }{4 \pi}\right)^2 \log \left( \frac{\Lambda}{\mu_{\textrm{\tiny{EW}}}} \right)   \, \left(C^{HL^{(1)}}_{\ell \ell} - C^{HL^{(3)}}_{\ell \ell} -C^{He}_{\ell \ell} - C^{Lu}_{\ell \ell 3 3} + C^{eu}_{\ell \ell 3 3} \right)\, ,
\end{eqnarray} 
where $\alpha_{e}\equiv e^2/(4 \pi)$, $e$ being the electric charge, and the overall normalization in the weak Hamiltonian follows the standard conventions adopted in refs.~\cite{Ciuchini:2015qxb,Ciuchini:2017mik,Ciuchini:2019usw}.

As anticipated in the Introduction, the set of operators of interest for the study of $R_{K^{(*)}}$ in eq.~\eqref{eq:SMEFT_matching_1loop} is also probed by EW precision data. Indeed, operators involving the Higgs field and lepton bilinears in the SMEFT induce modifications to EW-boson couplings that have been precisely measured at LEP/SLC, providing also an important test bed for lepton universality~\cite{Efrati:2015eaa,deBlas:2016ojx}. Modifications of the $Z$ couplings to the leptons can be induced also at loop level through the top-loop contribution~\cite{deBlas:2015aea}. In the leading-log approximation and at the leading order in the top Yukawa coupling, LUV effects can be generated by:
\begin{eqnarray}
\label{eq:OLuedRGE}
 \left.\Delta g_{Z,L}^{\ell\ell}\right|_{\mathrm{LUV}} & = &
 - \frac 12 \left( C_{\ell\ell}^{HL^{(1)}}+C_{\ell\ell}^{HL^{(3)}}\right)\frac{v^2}{\Lambda^2}-
 3 \left( \frac{ y_t \, v}{4 \pi \Lambda} \right)^2 \log\left(\frac{\Lambda}{\mu_{\textrm{\tiny{EW}}}} \right) \, C^{Lu}_{\ell\ell33}  \ , \\ \nonumber
 \left.\Delta g_{Z,R}^{\ell\ell}\right|_{\mathrm{LUV}} & = & 
 - \frac 12 C_{\ell\ell}^{He}\frac{v^2}{\Lambda^2}-
 3 \left( \frac{ y_t \, v}{4 \pi \Lambda} \right)^2 \log\left(\frac{\Lambda }{\mu_{\textrm{\tiny{EW}}}}\right) \, C^{eu}_{\ell\ell33} \ ,
\end{eqnarray}
where $\Delta g_{Z,L (R)}^{\ell\ell} \equiv g_{Z,L(R)}^{\ell\ell} - g_{Z,L (R)}^{\ell\ell,\textrm{SM}}$ is the deviation with respect to the left-handed (right-handed) leptonic couplings to the $Z$ boson in the SM theory.

Motivated by the previous observations, we would like to perform an EFT analysis of new physics models that can explain the flavour anomalies in the above-mentioned fashion, but exploring more generally the interplay of such SM extensions with EWPO. For that purpose, we consider an EFT analysis of new physics with the following assumptions:
\begin{itemize}
\setlength\itemsep{0em}
\item The solution to the flavour anomalies is obtained via radiative effects, such as those described in eq.~(\ref{eq:_C9_C10}).
\item Such NP can also contribute to EWPO at tree-level, in a flavour non-universal way.
\item Other effects that could enter in the previous observables via renormalization group (RG) mixing are either small or can be constrained better via other processes.
\end{itemize}
As we will see in \autoref{sec:UVtoymodels}, and can also be deduced using the results in \cite{deBlas:2017xtg}, it is not difficult to construct minimal BSM models where the previous conditions are satisfied.
From an EFT point of view, fulfilling these considerations requires the enlarging of the set of operators considered in eq.~\eqref{eq:SMEFT_op_HL} and also including the corresponding dimension-six interactions modifying the neutral and charged quark currents:
\begin{eqnarray} 
\label{eq:SMEFT_op_HQ}
O^{HQ^{(1)}}_{qq} &=& ( H^{\dagger} i \overset{\leftrightarrow}{D}_{\mu}H ) (\bar{Q}_{q} \gamma^{\mu}  Q_{q} )\,, \nonumber \\
O^{HQ^{(3)}}_{qq} &=& ( H^{\dagger} i \overset{\, \leftrightarrow_A}{D_{\mu}}H ) (\bar{Q}_{q} \gamma^{\mu} \tau^{A} Q_{q} )\,, \nonumber \\
O^{Hu}_{qq} &=& ( H^{\dagger} i \overset{\leftrightarrow}{D}_{\mu}H ) (\bar{u}_{q} \gamma^{\mu}  u_{q} )\,,\nonumber\\
O^{Hd}_{qq} &=& ( H^{\dagger} i \overset{\leftrightarrow}{D}_{\mu}H ) (\bar{d}_{q} \gamma^{\mu}  d_{q} )\,,
\end{eqnarray}
where $q=1,2,3$ identifies quark generations.\footnote{
In our SMEFT analysis we require these quark operators to be diagonal in a basis that is aligned, as much as possible, with the down-quark physical basis. This will be convenient to avoid possible dangerous tree-level FCNC effects~\cite{Silvestrini:2018dos}.
Similarly, we also assume lepton-flavour alignment with the charged-lepton mass basis.
}
In this regard, we note that EWPO cannot separate in a clean way contributions from the first family quarks, in particular in the $d$ sector. Therefore, and analogously to what was done in ref.~\cite{deBlas:2019wgy}, we identify deviations in the couplings of the EW bosons to the first and second family of the quarks via $C^{{HQ}^{(1,3)}}_{11} = C^{{HQ}^{(1,3)}}_{22} $, $C^{Hu}_{11} = C^{Hu}_{22} $, and $C^{Hd}_{11} = C^{Hd}_{22} $. This implicit $U(2)^3$ symmetry in the quark sector would in general also help to mitigate large contributions to FCNC. Note that, even in this situation, not all the Wilson coefficients related to eq.~\eqref{eq:SMEFT_op_HQ} can be well constrained with the EWPO. This is the case for the Wilson coefficient of $O^{Hu}_{33}$, which modifies the right-handed top quark coupling to the $Z$. This cannot be probed at tree level by $Z$-pole measurements. 

Introducing eq.~\eqref{eq:SMEFT_op_HQ} also modifies the EW couplings of the $Z$ to all fermions at the one-loop level, and in particular the leptonic couplings, $g_{Z,L(R)}^{\ell\ell}$. These are, however, flavour-universal effects. In our study, we propagate the leading $y_t$ effects of this kind, coming from the RG mixing with $O_{33}^{HQ^{(1)}}$. As we will see, given the comparatively weaker bound on the Wilson coefficient of that operator compared to the leptonic ones, these effects can be sizeable in the fit. It must be noted that, at the same order in the perturbative expansions we are considering, similar effects from $O_{33}^{Hu}$ could also have a non-negligible phenomenological impact. However, as explained before, $C_{33}^{Hu}$ cannot be directly bound in the EWPO fit. Hence, to avoid flat directions in our EFT analysis, we assume the RGE boundary condition $C_{33}^{Hu}=0$ to hold true. Excluding ${\cal O}_{33}^{Hu}$ and taking into account the aforementioned assumptions in the quark sector, eq.~\eqref{eq:SMEFT_op_HQ} adds a total of 7 new degrees of freedom into our EFT analysis.

Finally, for completeness, we also consider the effects of the four-lepton operator:
\begin{equation}
    \label{eq:SMEFT_op_LLLL}
    O_{1221}^{LL}=(\bar{L}_1 \gamma^\mu L_2) (\bar{L}_2 \gamma_\mu L_1) \ ,
\end{equation}
which contributes to the muon decay amplitude, and therefore alters the extraction of the value of the Fermi constant, $G_F$, which is one of the inputs of the SM EW sector.

The operators in eqs.~\eqref{eq:SMEFT_op_HL}, \eqref{eq:SMEFT_op_HQ} and \eqref{eq:SMEFT_op_LLLL}, with the assumptions mentioned before, saturate all the 17 degrees of freedom, i.e. combinations of operators, that can be constrained in a fit to EWPO in the dimension-six SMEFT framework~\footnote{In this regard, we should mention that at dimension six, in the Warsaw basis, EW observables are also affected by two more operators not discussed so far: ${ O}_{HWB}=(H^\dagger \tau^A H) W_{\mu\nu}^A B^{\mu\nu}$ and ${ O}_{HD}=\left|H^\dagger D_\mu H\right|^2$. Contrary to the set in eqs.~\eqref{eq:SMEFT_op_HL} and \eqref{eq:SMEFT_op_HQ}, these operators only induce oblique, and therefore flavour-universal, corrections in EW observables. Given our focus on LUV effects, we assume for ${O}_{HWB}$ and ${O}_{HD}$ that the corresponding Wilson coefficients are not generated by the NP at the scale $\Lambda$.}, while keeping flavour changing neutral currents in the light quark sector under control. Together with the 4 four-fermion operators from eq.~\eqref{eq:SMEFT_op_loop_lu}, this completes a total of 21 operators, which we include in the fit setup described in the next section. 

\section{Analysis strategy}
\label{sec:strategy}
We now proceed to discuss in more detail our EFT analysis. {Our aim is to pin down the picture that should address the present $B$ anomalies via one-loop SM RGE effects of flavour-conserving dimension-six operators, and respect at the same time the constraints from EW precision.} We can achieve this goal with a comprehensive global analysis that aims at combining EWPO and $b \to s \ell \ell$ data.\footnote{See ref.~\cite{Capdevila:2020rrl} for another recent analysis where $b \to s \ell \ell$ data and EW measurements have been combined, with the different scope of resolving tensions in the determination of the Cabibbo angle~\cite{Belfatto:2019swo,Grossman:2019bzp}.}

We perform a Bayesian analysis on the most recent set of $b \to s \ell \ell$ measurements together with the state-of-the-art theoretical information already implemented and described in ref.~\cite{Ciuchini:2019usw}. We include in our study EW physics following what originally done in ref.~\cite{Ciuchini:2013pca} and, more recently, in ref.~\cite{deBlas:2016ojx}. In particular, we adopt the list of observables reported in Table~1 of this reference, and allow for lepton non-universal contributions from heavy BSM physics in EWPO~\cite{Efrati:2015eaa,deBlas:2019wgy} within the framework described in \autoref{sec:theory}.

For this purpose we adopt the publicly available \HEPfit~\cite{deBlas:2019okz} package, a Markov Chain Monte Carlo (MCMC) framework built using the Bayesian Analysis Toolkit~\cite{2009CoPhC.180.2197C}.\footnote{All code and configuration files can be made available upon request.} In our analyses we vary $\mathcal{O}(100)$ parameters including nuisance parameters. The data that we use for the fits can be categorized as follows:
\begin{itemize}
\setlength\itemsep{0em}
    {\item The set of EWPO including the $Z$-pole measurements from LEP/SLD, the measurements of the $W$ properties at LEP-II, as well as several related inputs from the Tevatron and LHC measurements of the properties of the EW bosons~\cite{ALEPH:2005ab,Abe:2000uc,Group:2012gb,Schael:2013ita,Aaboud:2017svj,Khachatryan:2014iya,Abazov:2011ws}. The following lists the bulk of the EWPO included in the fits:
    \begin{gather*}
        M_H,~m_t,~\alpha_S(M_Z),~\Delta \alpha_{\mathrm{had}}^{(5)}(M_Z),\\
        M_{Z},~\Gamma_{Z},~R_{e,\mu,\tau},~\sigma_{\mathrm{had}}, ~A^{e,\mu,\tau}_{FB},~A_{e,\mu,\tau},~A_{e,\tau}(P_\tau),~ R_{c,b},~A^{c,b}_{FB},~A_{s,c,b},~R_{u+c}, \\
        M_{W},\Gamma_{W},~\mathrm{BR}_{W\to e \nu,\mu \nu,\tau \nu},~\Gamma_{W\to cs}/\Gamma_{W\to ud+cs},~\left|V_{tb}\right|;
    \end{gather*}
    }
    \item The angular distribution of $B\to K^{(*)}\ell^+\ell^-$ decays for both $\mu$ and $e$ final states in the large-recoil region.\footnote{We do not consider in this work low-recoil data, plagued by broad charmonium resonances, implying very large hadronic uncertainties. For analogous reasoning, we do not attempt to study here the baryon rare decay $\Lambda_{b} \to  \Lambda \, \mu^{+} \mu^{-}$ as well.} These include data from ATLAS~\cite{Aaboud:2018krd}, Belle~\cite{Wehle:2016yoi}, CMS~\cite{Khachatryan:2015isa,Sirunyan:2017dhj} and LHCb~\cite{Aaij:2015dea,Aaij:2020nrf}; we also include the branching fractions from LHCb~\cite{Aaij:2016flj}, and of $B\to K^*\gamma$\footnote{NP effects from dipole operators are strongly constrained as extensively investigated in ref.~\cite{Paul:2016urs}. However, radiative exclusive $B$ decays still provide relevant information about hadronic effects~\cite{Ciuchini:2018anp}.} for which we use the HFLAV average~\cite{Amhis:2019ckw};
    \item Branching ratios for $B^{(+)} \to K^{(+)} \mu^+\mu^-$ decays in the large-recoil region measured by LHCb~\cite{Aaij:2014pli};
    \item The angular distribution of $B_s\to \phi\mu^+\mu^-$~\cite{Aaij:2015esa} and the branching ratio of  the decay $B_s\to\phi\gamma$~\cite{Aaij:2012ita}, measured by LHCb;
    \item The lepton universality violating ratios $R_K$~\cite{Aaij:2019wad} and $R_{K^*}$~\cite{Aaij:2017vbb} from LHCb and Belle~\cite{Abdesselam:2019wac};
    \item Branching ratio of $B_{(s)}\to \mu^+\mu^-$ measured by LHCb~\cite{Aaij:2017vad}, CMS~\cite{Chatrchyan:2013bka}, and ATLAS~\cite{Aaboud:2018mst}; we also use the upper limit on $B_s\to e^+e^-$ decay reported recently by LHCb~\cite{Aaij:2020nol}. 
\end{itemize}

For the $B\to K^*\ell^+\ell^-$ channel, as in previous works~\cite{Ciuchini:2016weo,Ciuchini:2017mik,Ciuchini:2017gva,Ciuchini:2018xll,Ciuchini:2018anp,Ciuchini:2019usw}, we consider two different scenarios for hadronic contributions stemming from long-distance effects~\cite{Khodjamirian:2010vf,Jager:2014rwa,Lyon:2014hpa}. We take into account a conservative approach (Phenomenological Data Driven or PDD) as originally proposed in \cite{Ciuchini:2015qxb}, and refined in ref.~\cite{Ciuchini:2018anp}, and a more optimistic approach based on the results in~\cite{Khodjamirian:2010vf} (Phenomenological Model Driven or PMD). For the PDD model, a quite generic model of hadronic contributions is simultaneously fitted to $b \to s \ell \ell$ data together with the effects coming from NP. Within this approach, a net assessment of the presence of BSM physics is only possible via observables sensitive to LUV effects. See the discussion in ref.~\cite{Ciuchini:2019usw} for more details. For the PMD approach we use the dispersion relations specified in~\cite{Khodjamirian:2010vf} to constrain the hadronic contributions in the entire large-recoil region considered in the analysis. This leads to much smaller hadronic effects in the~$B\to K^*\ell^+\ell^-$ amplitudes~\cite{Ciuchini:2016weo}, which significantly affects NP results of global analysis~\cite{Ciuchini:2019usw}.

We have characterized our study by considering several different scenarios for the SMEFT fit. In particular, we would like to clarify the sets of data and operators used in each of these fit scenarios, which are organized as follows:

\begin{itemize}
\setlength\itemsep{0em}
    \item {\bf EW}:
    In this fit we simultaneously vary the Wilson coefficients of the {\it 17 operators} in eqs.~\eqref{eq:SMEFT_op_HL},~\eqref{eq:SMEFT_op_HQ}, and \eqref{eq:SMEFT_op_LLLL}, as presented in \autoref{sec:theory}. This fit includes EW precision measurements only, and it is performed under the assumptions listed in \autoref{sec:theory}.
    \item  {\bf EW (SL-4F Only)}: This refers to a fit done with the Wilson coefficients of the {\em SL-4F operators} involving the right-handed top current, reported in eq.~\eqref{eq:SMEFT_op_loop_lu}. This scenario incorporates the assumption that BSM enters the modifications of the $Z$ couplings to muons and electrons through top-quark loops only.
    \item {\bf EW \& Flavour}: In these fits we vary the Wilson coefficients of all the {\em 21 operators} given in eq.~\eqref{eq:SMEFT_op_HL},~\eqref{eq:SMEFT_op_HQ}, and eq.~\eqref{eq:SMEFT_op_LLLL}, together with eq.~\eqref{eq:SMEFT_op_loop_lu}.
    We use all the EW data and include all the flavour observables listed at the beginning of this section. This scenario comes in two varieties, PDD and PMD, as explained above.
    \item {\bf Flavour}: These fits exclusively include the Wilson coefficients of the {\em 4 operators} (both electrons and muons) appearing in eq.~\eqref{eq:SMEFT_op_loop_lu}, and are done including only flavour data, i.e. excluding EW measurements. Results are again distinguished for the PDD and PMD cases. 
\end{itemize}

\section{Results from the SMEFT}
\label{sec:EFT_results}

\subsection[Analysis of EW and \texorpdfstring{$b \to s \ell \ell$}{b to s ll} data]
{Analysis of EW and \boldmath$b \to s \ell \ell$ data}
\label{sec:GEN_EFT_results}

As a first step in our analysis, we reproduced the outcome of the EW fit originally obtained in ref.~\cite{Efrati:2015eaa} using \HEPfit. 
Then, we expanded upon the standard EW results through the study of the {\bf EW} scenario introduced in the previous section, yielding constraints on the Wilson coefficients of the SMEFT operators involving, in particular, dimension-six operators with a Higgs-doublet current, and including also leading-loop effects under the working hypotheses stated in \autoref{sec:theory}. The subset of these operators containing leptonic currents can give rise to non-universal modifications of EW gauge-boson couplings. Assuming NP integrated out at the heavy scale $\Lambda > v$, these operators also contribute via RGE flow to $b \to s \ell \ell $ observables at one loop, see eq.~\eqref{eq:SMEFT_matching_1loop}.

\begin{sidewaysfigure}[ht]
    \centering
    \includegraphics[width=\textwidth]{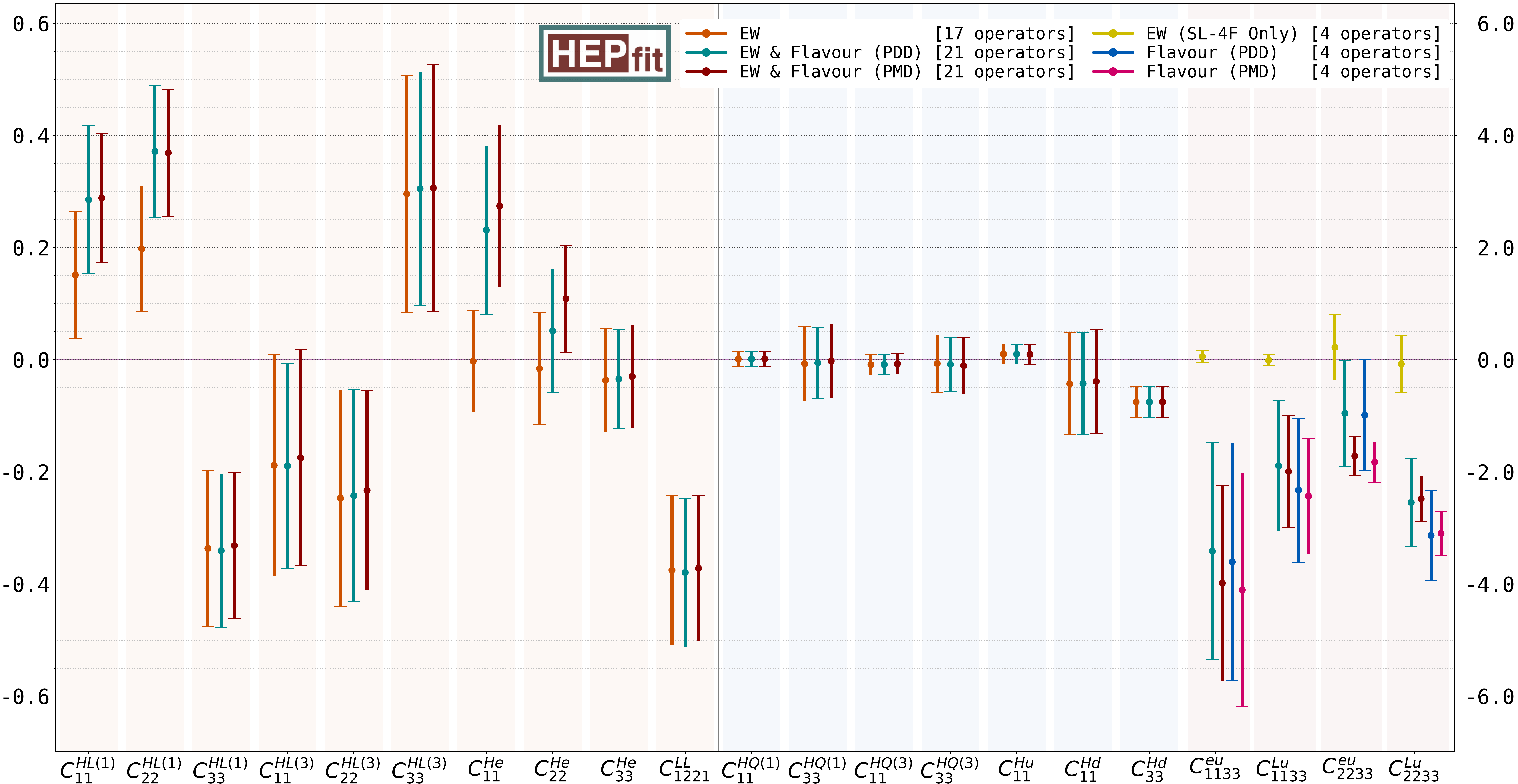}
    \caption{Mean and standard deviation of the marginalized posterior distributions for each of the Wilson coefficients (in TeV$^{-2}$) considered in the different fits described in \autoref{sec:strategy}.
    Note that each fit assumes a different set of non-zero operators:
    EW -- 17 operators presented in eqs.~\eqref{eq:SMEFT_op_HL}, \eqref{eq:SMEFT_op_HQ} and \eqref{eq:SMEFT_op_LLLL};  EW(SL-4F Only) -- four-fermion operators in eq.\eqref{eq:SMEFT_op_loop_lu}; Flavour (PDD) and (PMD) are the fits with the operators in eq.\eqref{eq:SMEFT_op_loop_lu}, where (PDD) and (PMD) refer to the various assumptions on the hadronic long-distance effects in the flavour sector; EW \& Flavour (PDD) and (PMD) stand for the fits including the 21 operators in eqs.~\eqref{eq:SMEFT_op_HL}, \eqref{eq:SMEFT_op_loop_lu}, \eqref{eq:SMEFT_op_HQ} and \eqref{eq:SMEFT_op_LLLL}. (Note the different scaling in the axes quantifying the size of the bounds presented in each half of the figure.)
    }
    \label{fig:ew_flav_bounds}
\end{sidewaysfigure}

\begin{figure}[ht!]
    \centering
    \includegraphics[width=\textwidth]{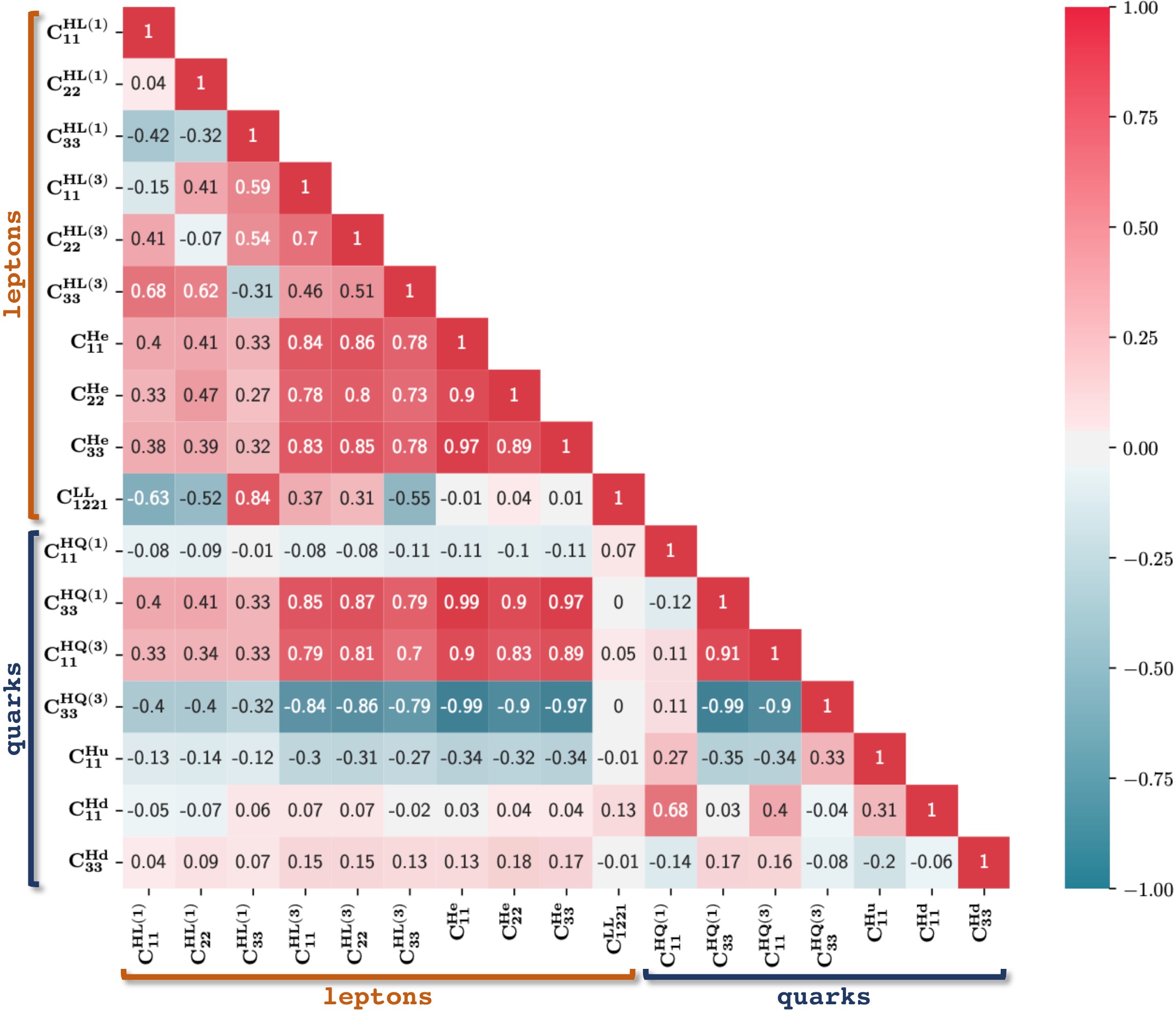}
    \caption{\it The correlation matrix extracted from the SMEFT analysis of the set of independent operators in eqs.~\eqref{eq:SMEFT_op_HL}, \eqref{eq:SMEFT_op_HQ}, \eqref{eq:SMEFT_op_LLLL} in the \textbf{EW} scenario introduced in \autoref{sec:strategy}. The two distinct groups of Wilson coefficients associated to leptonic and quark interactions are remarked as ``leptons'' and ``quarks'', respectively.}
    \label{fig:ew_corr}
\end{figure}

{On the left side of \autoref{fig:ew_flav_bounds}, we show in orange the bounds from the {\bf EW} fit on the Wilson coefficients of the operators with leptonic currents in terms of mean and standard deviation of the marginalized posterior probability density function. We observe compatibility with the SM within the 2$\sigma$ level. Note that EW data strongly correlate the operators under consideration among themselves, as can be seen in the correlation matrix presented in \autoref{fig:ew_corr}.\unskip\parfillskip 0pt \par}

\FloatBarrier

\noindent where away from the photon pole, $R_{K^{(*)}}^{\textrm{\tiny SM}}$ are predicted to be unity at percent level~\cite{Bordone:2016gaq}.

\begin{figure}[htp!]
    \centering
    \includegraphics[width=\textwidth]{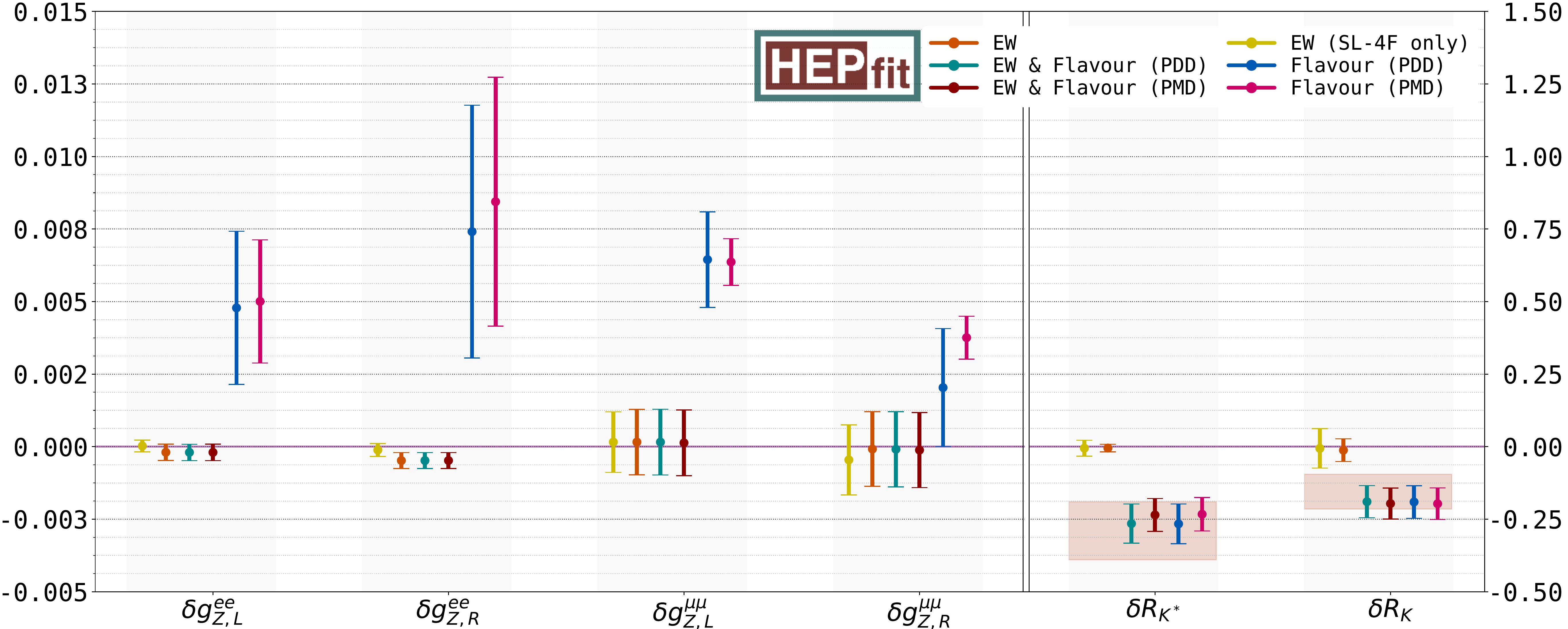}
    \caption{\it Mean and standard deviation of the marginalized posterior of the key set of observables for this work, in relation to the tension between $b \to s \ell \ell$ anomalies and LEP/SLD measurements. In particular, the left panel shows the deviations in the effective $Z\ell\ell$ couplings, normalized by SM values. The right panel, on the other hand, shows the deviation from the nominal SM values of the lepton universality violating ratios, see eq.~\eqref{eq:SMdev}, with the red boxes indicating the region selected by the experimental measurements of $R_{K,(K^*)}$.
   }
    \label{fig:ew_flav_dg}
\end{figure}

\noindent In particular, the strong correlation between the operators with quarks and leptons is introduced by the non-negligible one-loop universal contribution of the operator ${\cal O}_{33}^{HQ^{(1)}}$ to all the EW couplings, as anticipated at the end of \autoref{sec:EFT_results}. 
With the direct bound on $C_{33}^{HQ^{(1)}}$ being relatively weak compared to the limits on the leptonic operators, such effects in the leptonic couplings can be sizable. 

This leads to a relaxation of the naive bounds on $C^{HL^{(1)}}_{\ell\ell}$, $C^{HL^{(3)}}_{\ell\ell}$ and $C^{He}_{\ell\ell}$ that one would obtain in a tree-level analysis.
To illustrate this, we present in \autoref{app:EW} a comparison with the results from such a tree level analysis of the EW fit. The results in \autoref{fig:ew_corr} can then be compared to those in \autoref{fig:ew_corr_tree} where, as it is apparent, there is a substantial decoupling between the dimension-six operators made of Higgs doublets and quark bilinears from the leptonic ones.

The impact of these operators on the key observables for the present discussion is reported in \autoref{fig:ew_flav_dg}. There, we collect mean and standard deviation on the shift in the $Z$ coupling to light leptons (normalized to the corresponding SM value),and on the effect on $R_{K^{(*)}}$ in the dilepton-mass range $[1.0,6.0]$~GeV$^2$: 
\begin{equation}
\label{eq:SMdev}
    \delta g_{Z,L(R)}^{ee(\mu\mu)} \equiv g_{Z,L(R)}^{ee(\mu\mu)}{\big/}g_{Z,L(R)}^{ee(\mu\mu),\textrm{\tiny SM}}-1  \ , \ \delta R_{K^{(*)}} \equiv R_{K^{(*)}}^{ } - R_{K^{(*)}}^{\textrm{\tiny SM}} \ ,
\end{equation}

Note that EW measurements tightly constrain NP effects modifying the 
EW gauge boson couplings to electrons, and also forbid deviations beyond the per-mille level in the case of  couplings to muons. This translates into strong bounds on the Wilson coefficients $C^{H L^{(1,3)},H e}_{\ell \ell}$. Hence, the one-loop contribution to $R_{K^{(*)}}$ from $O^{H L^{(1,3)},H e}_{\ell \ell}$ comes out to be tiny. We can then move our attention to the {\bf EW (SL-4F Only)} scenario, reported in yellow in \autoref{fig:ew_flav_bounds} and \autoref{fig:ew_flav_dg}, and find a similar conclusion. Indeed, EW data once again strongly constrain the NP Wilson coefficients related to $O^{eu,L u}_{\ell \ell 33}$ -- the SL-4F operators -- implying all the four NP Wilson coefficients to be compatible with 0. However, note that unlike the previous case, $C^{Lu,eu}_{\ell \ell 3 3}$ only contribute at one loop to $\delta g_{Z,L(R)}^{\ell\ell}$ and $\delta R_{K^{(*)}}$ in eq.~\eqref{eq:SMdev}. Consequently, the resulting impact on $b \to s  \ell \ell$ flavour observables can be larger than the one in the {\bf EW} scenario. As depicted in \autoref{fig:ew_flav_dg}, however, there is still an overall tension between EWPO bounds (in yellow) and the experimental measurements of $R_K$ and $R_{K^*}$ (indicated by the shaded red boxes in the right side of the figure) at the 3$\sigma$ level.

To frame this tension from a different perspective, let us now focus on the set of flavour measurements as previously done in ref.~\cite{Ciuchini:2019usw}. In \autoref{fig:ew_flav_bounds} we also show the constraints on the four Wilson coefficients of eq.~\eqref{eq:SMEFT_op_loop_lu} coming from $b \to s\ell\ell$ data, in what we dubbed as the {\bf Flavour} scenario. We present the PMD case, corresponding to an optimistic approach to QCD power corrections, in pink, while the more conservative PDD case is shown in blue. We observe that in both cases a muonic solution to $B$ anomalies stands out, with $C^{Lu}_{2233}$ different from 0 at more than 3$\sigma$ in the PDD case, and at roughly 6$\sigma$ in the PMD one.

We stress that the difference between the results obtained in the PMD and in the PDD case is substantially driven by the angular analysis of $B \to K^{*} \mu \mu$. In particular, only within the PDD approach the fully left-handed solution to $B$ anomalies, $C_{9, \ell} = - C_{10, \ell}$, is favoured by data (signalled here by the Wilson coefficient of $O^{eu}_{\ell \ell 3 3}$ being compatible with 0 at 1$\sigma$, see the results in blue in \autoref{fig:ew_flav_bounds}). In addition, an electron resolution of $B$ anomalies is, once again, viable only within PDD~\cite{Ciuchini:2017mik,Ciuchini:2019usw}.

In the {\bf Flavour} scenario one can also predict the induced shift in the $Z$-boson couplings according to eq.~\eqref{eq:OLuedRGE}, and these are shown in \autoref{fig:ew_flav_dg}. As can be seen, $\delta g^{\ell \ell}_{Z,L,R}$ would receive large contributions at one loop from $O^{Lu,eu}_{\ell \ell 33}$ in correspondence to the one-loop MFV-like resolution of $B$ anomalies. Such contribution would be, however, now in tension with the results from EW precision tests. In particular, as a reflection of the main role played by $O^{Lu}_{2233}$ in the {\bf Flavour} fit to the four NP Wilson coefficients considered, $g^{\mu \mu}_{Z,L}$ shows the most important deviation from the SM value. Also, the prediction of $g^{ \mu \mu}_{Z,L(R)}$ becomes indirectly sensitive to the underlying treatment of hadronic uncertainties adopted for the study of $b \to s $ data. Therefore, we observe that within the PMD approach, the inconsistency between what is needed to address $B$ anomalies and what is required by EW measurements is even more severe than the 3$\sigma$ established in the {\bf EW (SL-4F Only)} scenario, and imprinted also in the {\bf Flavour} fit with the PDD approach. In fact, we stress once again that adopting light-cone sum-rule results~\cite{Khodjamirian:2010vf} for the long-distant effects in $B \to K^{*} \ell \ell$ decay, the tension between $B$ anomalies and EW data reaches the 6$\sigma$ level.

\begin{figure}[t]
    \centering
    \includegraphics[width=0.9\textwidth]{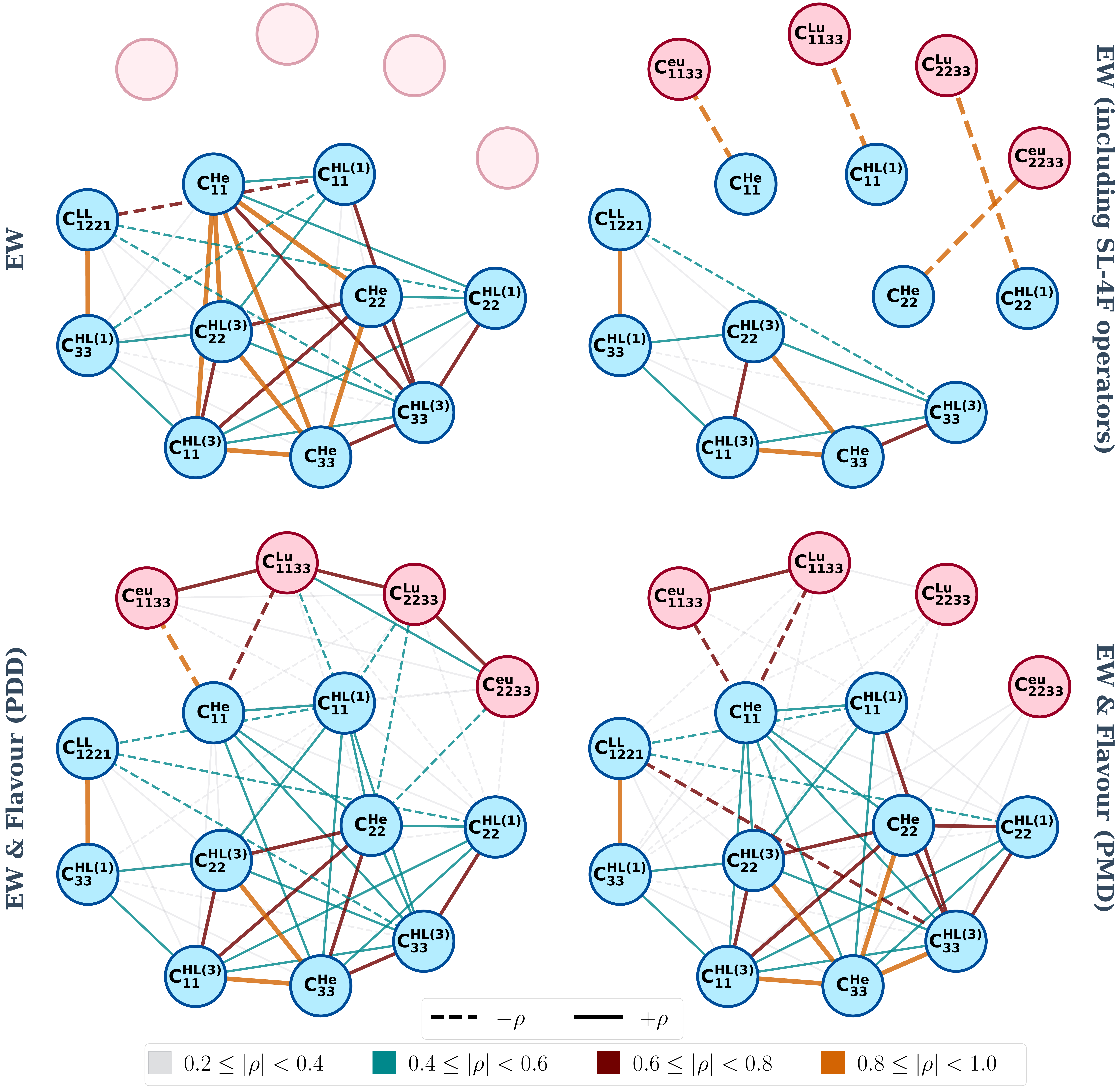}
    \caption{\it Correlations among dimension-six operators involving leptonic currents in different scenarios. In the upper side we show the \textbf{EW} fit (upper-left panel), and the scenario where in the same setup the SL-4F operators are also included (upper-right panel), highlighting the anti-correlation among the set of Wilson coefficients  $C^{HL^{(1)}}_{\ell\ell}, C^{He}_{\ell\ell}$ and $C^{Lu,eu}_{\ell\ell33}$.
    In the lower-side panels we show how $b \to s \ell \ell$ measurements break these degeneracies, showing the \textbf{Flavour} fit for the PDD case (lower-left panel), and the PMD one (lower-right panel). 
    }
    \label{fig:ew_flav_corr}
\end{figure}

So, how do we reach a consensus between $b \to s \ell \ell$ measurements and EWPO?

\noindent Succinctly, an obvious solution which satisfies these constraints is a class of models where $R_{K^{(*)}}$ anomalies are addressed at tree level and where modifications to $Z$-lepton-lepton vertices are at the same time suppressed. However, these models would not offer a solution to $B$ anomalies of the MFV type envisaged so far, namely they would rely on the existence of sizeable new sources of flavour violation. At this point, we would like to emphasize that a combined fit of EW and flavour observables offers a new insight into this matter: it highlights strong correlations between the dimension-six operators $O^{Lu(eu)}_{\ell\ell33}$ and $O^{HL^{(1)}(He)}_{\ell\ell}$ as is evident from \autoref{fig:ew_flav_corr}. This figure presents a pictorial representation of the correlations between the leptonic operators included in the different fits. 

Apart from the fits introduced in the previous section, for illustration purposes we also show in \autoref{fig:ew_flav_corr} the correlations obtained in a variant of the ${\bf EW}$ fit including also the four-fermion operators $O^{Lu(eu)}_{\ell\ell33}$, labelled as {\bf EW (including SL-4F operators)}. 
This is shown in the upper-right corner of the figure. As can be seen in that panel, and one could deduce from the relations in eq.~\eqref{eq:OLuedRGE}, in a pure EW fit adding the four-fermion operators would simply introduce 4 flat directions. These are illustrated by the links connecting the $C^{eu}_{\ell\ell 33}$ ($C^{Lu}_{\ell\ell 33}$) and $C^{He}_{\ell\ell}$ ($C^{HL^{(1)}}_{\ell\ell}$) operators, corresponding to 100\% anti-correlation.
Such flat directions are lifted upon the introduction of the flavour measurements of $R_{K}$ and $R_{K^*}$, as can be seen in the lower panels of \autoref{fig:ew_flav_corr} for the {\bf EW \& Flavour} fits.
Even then, due again to relations in eq.~\eqref{eq:SMEFT_matching_1loop} and~\eqref{eq:OLuedRGE} and the comparatively different precision of the EW and flavour measurements, sizable correlations remain. 

In \autoref{fig:ew_flav_bounds} the imprint of these correlations is a shift of central values and an increase on the bounds on the corresponding Wilson coefficients, with red and green bars representing the outcome of the fit in the {\bf EW \& Flavour} scenario within the {\bf PMD} and {\bf PDD} approaches, respectively. The interplay between $O^{Lu(eu)}_{\ell\ell33}$ and $O^{HL^{(1)}(He)}_{\ell\ell}$ is evident when comparing the reported red and green bounds versus the orange EW constraints on $C^{HL^{(1)}(He)}_{\ell\ell}$, and the yellow ones for $C^{Lu(eu)}_{\ell\ell33}$. Consequently, as clearly depicted in \autoref{fig:ew_flav_dg}, looking at the red and green ranges reported for the {\bf EW \& Flavour} scenario, $R_{K^{(*)}}$ puzzles are solved with EW precision being respected. It is important to emphasize that, despite the significant correlation between quark and lepton operators introduced by the one-loop effects of $C_{33}^{HQ^{(1)}}$, quark operators play no significant role in reconciling the EWPO constraints with the solution to $B$ anomalies. This will become clearer in the next section, but can be easily understood from the fact that, as mentioned before, quark and lepton constraints are somewhat uncorrelated in the tree-level EW fit, and the fact that the one-loop corrections effect induced by $C_{33}^{HQ^{(1)}}$ are flavour universal.

\subsection{A minimal EFT picture}
Finally, let us draw what would be the minimal picture for NP out of the general analysis obtained with the 21 operators considered in the {\bf EW \& Flavour} scenario. Indeed, a simpler picture will serve as a guideline for the UV models discussed in \autoref{sec:UVtoymodels}. 
As mentioned before, given the hadronic uncertainties at hand, the most economic explanation addressing in particular $R_{K^{(*)}}$ anomalies resides in the NP contribution from the fully left-handed operator, $O^{LQ}_{\ell \ell 2 3}$. In the present context this operator is generated at one loop by $O^{Lu}_{\ell \ell 3 3}$, according to eq.~\eqref{eq:SMEFT_matching_1loop}.

\begin{figure}[tp]
    \includegraphics[width=\textwidth]{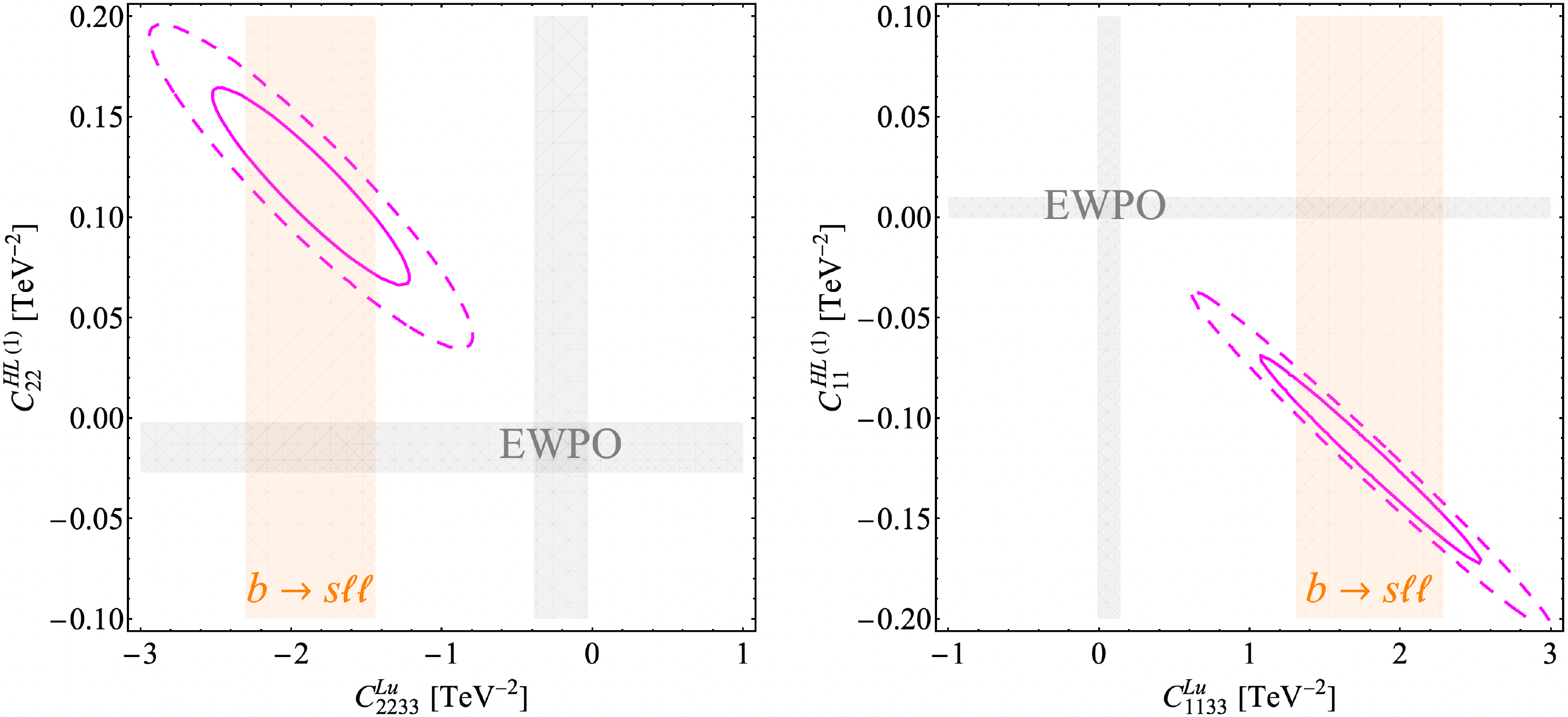}
    \caption{\it The most economic EFT picture where $B$ anomalies can be reconciled at one loop with EWPO. In (dashed) magenta the 1(2)$\sigma$ correlation between the Wilson coefficients of the operators responsible of addressing $B$ anomalies without any source of flavour violation beyond the Yukawa couplings of the SM. The minimal scenario involves LUV effects in the (electron) muon sector as highlighted by the 1$\sigma$ orange band in the (right) left panel, originated from $b \to s \ell \ell$ data analyzed with a conservative approach to hadronic uncertainties. In same figure, the 1$\sigma$ region allowed by EWPO within a single-operator analysis, horizontal and vertical grey bands.}    
    \label{fig:2D_correlations}
\end{figure}

Then, in \autoref{fig:2D_correlations} we show in orange the overall constraint from $b \to s \ell \ell$ data on $C^{Lu}_{\ell \ell 3 3}$ within the most conservative approach to long-distance effects, i.e. the PDD one. In particular, in the left (right) panel we report the constraint on the muonic (electronic) scenario. In the same figure, we highlight with the vertical gray band the bound derived from the full correlated set of EWPO on the same operator. From the comparison of the orange and gray single-operator bounds, the tension between flavour and EW measurements is manifest at the 3$\sigma$ level in the left panel of \autoref{fig:2D_correlations}. It gets even more pronounced in the right panel due to the precise probe of NP that EW gauge-boson couplings to electrons provide. In the same \autoref{fig:2D_correlations}, we also show with the horizontal gray band the result of the EWPO constraints applied this time on the NP contribution coming exclusively from the operator $C^{HL^{(1)}}_{\ell \ell}$. Note that this operator would also contribute to $R_{K^{(*)}}$ at one loop, but the size needed would be $\mathcal{O}(1)$ and it is out of scale in the vertical axis of the plot.

Most importantly, in the same figure we display in (dashed) magenta the 1(2)$\sigma$ contour where EW data are reconciled with the one-loop MFV explanation of $B$ anomalies when a combined fit of the NP contributions from these two operators is performed. {Therefore, heavy BSM degrees of freedom that, once integrated out, generate sizeable contributions both  to the Wilson coefficient of $O^{HL^{(1)}}_{\ell \ell}$ and of $C^{Lu}_{\ell \ell33}$ are the key aspect of this scenario that addresses $B$ anomalies without requiring sources of flavour violation beyond SM ones.}

Finally, note that the role played here by $O^{Lu}_{\ell \ell 3 3}$ could be shared, in part, with $O^{eu}_{\ell \ell 3 3}$, depending on how much departure is actually required from the fully left-handed solution to $B$ anomalies. As already noted, this fact critically depends on the information stemming from $B \to K^{*} \mu \mu$~\cite{Ciuchini:2019usw}. On general grounds, to relieve the bounds from EWPO, the presence of $O^{eu}_{\ell \ell 3 3}$ would also necessitate sizeable NP effects from $O^{He}_{\ell \ell}$. 

As a last comment of this section we would also like to highlight that in the class of models considered the prediction for the LUV observable $R_{K}$ is always close to the one for $R_{K^{*}}$: any hint of NP coming from $R_{K^{*}}/R_{K} \neq 1$~\cite{Hiller:2014ula,Hurth:2014vma,Hiller:2014yaa,Hiller:2017bzc} would not be addressed within the NP models considered here, mainly involving the operators in eq.~\eqref{eq:SMEFT_op_HL} and~\eqref{eq:SMEFT_op_loop_lu}. In the following sections we will put our focus on the economic EFT scenario captured in \autoref{fig:2D_correlations} to build up simple UV scenarios realizing the EFT picture here delineated.

\section{Directions for UV models}
\label{sec:UVtoymodels}

In this section we discuss how the lesson derived from the SMEFT picture illustrated, in particular, in \autoref{fig:2D_correlations}, can be realized in a minimal extension of the SM. Here, we explicitly show how models involving a new $Z^\prime$ gauge boson around the TeV scale provide the most economic example of the correlations advertised in the previous section. 
This can be achieved if we have a $Z^\prime$ coupled both to top and lepton SM fields.
These couplings can be obtained introducing vector-like top and muon/electron partners reasonably close to the EW scale~\cite{Kamenik:2017tnu,Fox:2018ldq}, making this class of models potentially interesting also from the point of view of naturalness in the Higgs sector.
Finally, we will also briefly comment on possible alternative scenarios that can be obtained with leptoquarks.

\subsection{\texorpdfstring{Z$^{\prime}$}{Z'} with vector-like partners}
\label{sec:mod_Zprime}

Let us start with the baseline presented originally in ref.~\cite{Kamenik:2017tnu}. A simple extension of the SM, able to address $B$ anomalies, and  that does not introduce any explicit new source of flavour violation, can be conceived as follows:
\begin{itemize}
    \item The SM gauge group, $SU(3)_{c} \otimes SU(2)_{L} \otimes U(1)_{Y}$, is extended by a new Abelian gauge group, $U(1)_{X}$, under which SM fields are neutral;
    \item There is a new complex scalar field $\mathcal{S}$ that spontaneously breaks $U(1)_{X}$, giving a mass to the gauge boson $X_{\mu}$ equal to $m_{Z^{\prime}} = g_{X} \langle \mathcal{S} \rangle $;
    \item A coloured vector-like top partner, $\mathcal{T}$, properly charged under $U(1)_{X}$ and $U(1)_{Y}$ can mix with the right-handed top-quark field $u_{3}$ via a Yukawa interaction with $\mathcal{S}$; 
    \item A vector-like muonic partner, $\mathcal{M}$, doublet of $SU(2)_{L}$ and charged under $U(1)_{X,Y}$, can mix with the muonic doublet $L_2$ via another Yukawa coupling of $\mathcal{S}$;
    \item The couplings controlling the kinetic-mixing term, $X_{\mu \nu} B^{\mu \nu}$, and the quadratic scalar mixing, $\mathcal{S}^{\dagger}\mathcal{S} H^{\dagger} H$, are set to be phenomenologically negligible.\footnote{Using naive dimensional analysis, both kinetic and scalar quadratic mixing should appear beyond the tree level suppressed at least by a loop factor and the corresponding SM-partner rotation angles.} 
\end{itemize}

Then, the UV model is completely characterized by eight new parameters: the gauge coupling $g_{X}$, the mass $\mu_{\mathcal{S}}$ and quartic $\lambda_{\mathcal{S}}$ of the renormalizable potential of $\mathcal{S}$, the new Yukawa couplings $Y_{\mathcal{T},\mathcal{M}}$, here taken to be real, and the vector-like mass-term parameters $M_{\mathcal{T},\mathcal{M}}$.
In particular, the Lagrangian of the model contains the following terms:
\begin{equation}
    \label{eq:new_Yukawa_Mu}
    M_{\mathcal{T}} \bar{\mathcal{T}_R} \mathcal{T}_L + M_{\mathcal{M}} \bar{\mathcal{M}}_R \mathcal{M}_L +
    Y_{t} \bar{u}_{3} \tilde{H}^\dagger Q_{3}  
    + Y_{\mathcal{T}} \bar{u}_{3} \mathcal{T}_L \mathcal{S} 
    + Y_{\mu} \bar{e}_{2}  H^\dagger  L_{2}
    + Y_{\mathcal{M}}\bar{\mathcal{M}}_R L_{2} \mathcal{S} + \mathrm{h.c.}  \ , 
\end{equation}
that characterize the mixing pattern of SM fields and vector-like partners.\footnote{Note that upon an opposite $U(1)_{X}$ charge assignment for the vector-like fermionic partners than the one implicitly assumed, one should replace in eq.~\eqref{eq:new_Yukawa_Mu} $\mathcal{S}$ with $S^{\dagger}$.}
 Symmetry breaking of $U(1)_X$ is triggered by $\langle \mathcal{S} \rangle^2 = -\mu^2_{\mathcal{S}}/(2 \lambda_{\mathcal{S}}) \equiv \eta^2 \neq 0$, that implies the following fermionic mixing patterns:
\begin{eqnarray}
\label{eq:Mixing_Partner}
& \textrm{top sector:} & \ 
\left(  \begin{array}{cc}
\bar{u}_3 & \overline{\mathcal{T}}_R
\end{array} \right) \, \begin{pmatrix}
\frac{Y_t  \, v}{\sqrt{2}} \ &  \frac{Y_{\mathcal{T}} \eta}{\sqrt{2}} \ \\
0 \ & \ M_\mathcal{T}
\end{pmatrix} \,
\left(  \begin{array}{c}
U_3 \\  \mathcal{T}_L
\end{array} \right) \ + \ \mathrm{h.c.} \, , \\
& \textrm{muon sector:} & \ 
\left(  \begin{array}{cc}
\bar{e}_2 & \overline{\mathcal{M}}_{R}
\end{array} \right) \, \begin{pmatrix}
\frac{Y_\mu v}{\sqrt{2}} \ & 0\ \\
\frac{Y_{\mathcal{M}} \eta }{ \sqrt{2}}  \ & \ M_\mathcal{M}
\end{pmatrix} \,
\left(  \begin{array}{c}
E_{2} \\  \mathcal{M}_L
\end{array} \right) \ + \ \mathrm{h.c.} \, , \nonumber
\end{eqnarray}
where $U_{i}$ ($E_{i}$) indicates the $Q_{i}$-component ($L_{i}$-component) with weak isospin $1/2$ (-1/2). Using the determinant and trace of the squared mass matrices, one can easily show that the eigenvalues $m_{t,\mathcal{T}}$ and $m_{\mu,\mathcal{M}}$ must satisfy~\cite{Kamenik:2017tnu}:
\begin{eqnarray}
    \label{eq:Mass_Eigenvalues}
     m_{t,\mu} \,  m_{\mathcal{T,M}} & = & \frac{1}{\sqrt{2}} Y_{t,\mu} v M_{\mathcal{T,M}} \ , \\ m_{t,\mu}^2 + m_{\mathcal{T,M}}^2 & = & 
     M_{\mathcal{T,M}}^2 + \frac{1}{2} (Y_{t,\mu} \, v)^2 + \frac{1}{2} (Y_{\mathcal{T,M}} \, \eta)^2 \ , \nonumber
\end{eqnarray}
that in the decoupling limit clearly yield: $m_{t,\mu} \simeq Y_{t,\mu} v / \sqrt{2}$, $m_{\mathcal{T,M}} \simeq M_{\mathcal{T,M}}$.

Defining for the top sector the rotation matrix from the interaction to the mass basis following the convention:

\begin{equation}
    \label{eq:rot_phys_states} 
    \left(  \begin{array}{c}
t_{R (L)} \\  \mathcal{T}^{\prime}_{R (L)}
\end{array} \right) = \,
\begin{pmatrix}
\cos \theta^{t}_{R(L)} \ & \ - \sin \theta^{t}_{R(L)} \ \\
\sin \theta^{t}_{R(L)}  \ & \ \ \ \, \cos \theta^{t}_{R(L)}
\end{pmatrix} \,
    \left(  \begin{array}{c}
u_{3} (U_{3}) \\  
\mathcal{T}_{R (L)}
\end{array} \right) \ ,
\end{equation}
and doing similarly for the muonic sector, the mixing angles between SM fields, $t$ and $\mu$,  and their partner mass eigenstates, $\mathcal{T}^{\prime}$ and $\mathcal{M}^{\prime}$, can be conveniently expressed in terms of the dimensionless ratios $\xi_{\mathcal{T,M}}$ and $\varepsilon_{t,\mu}\, $:
\begin{eqnarray}
    \label{eq:partner_mixing}
    & \  \tan 2 \theta_{R}^{t} = \frac{2 \xi_{\mathcal{T}}}{\xi_{\mathcal{T}}^2-\varepsilon_{t}^2-1} \, , \, 
    \ \, \tan 2 \theta_{L}^{t} = \frac{2 \varepsilon_{t}}{\xi_{\mathcal{T}}^2-\varepsilon_{t}^2 +1} \, , \,   \textrm{with} \ \varepsilon_{t} \equiv \frac{Y_{t} v}{Y_{\mathcal{T}} \eta},~\xi_{\mathcal{T}} \equiv \frac{\sqrt{2} M_{\mathcal{T}}}{\eta Y_{\mathcal{T}}} \, ; \ \\
    & \  \tan 2 \theta_{R}^{\mu} = \frac{2 \varepsilon_{\mu}}{\xi_{\mathcal{M}}^2-\varepsilon_{\mu}^2+1} \, , \, 
    \tan 2 \theta_{L}^{\mu} = \frac{2 \xi_{\mathcal{M}}}
    {\xi_{\mathcal{M}}^2-\varepsilon_{\mu}^2 -1}  \, , \, 
    \textrm{with} \ \varepsilon_{\mu} \equiv \frac{Y_{\mu} v}{Y_{\mathcal{M}} \eta},~\xi_{\mathcal{M}} \equiv \frac{\sqrt{2} M_{\mathcal{M}}}{\eta Y_{\mathcal{M}}} \, .
    \nonumber
\end{eqnarray}
In a perturbative expansion in $\varepsilon_{t,\mu}$, eq.~\eqref{eq:partner_mixing} clearly shows that the mixing in the top sector proceeds mainly through $\tan\theta^{t}_{R} \simeq 1/\xi_{\mathcal{T}}$, while in the muonic sector one has  $\tan\theta^{\mu}_{L} \simeq 1/\xi_{\mathcal{M}}$ and very tiny $\tan\theta^{\mu}_{R}$.

Hence, for $\varepsilon_{t,\mu}/\xi_{\mathcal{T,M}}= Y_{t,\mu} v/\sqrt{2} M_{\mathcal{T,M}} < 1$, the leading couplings of the $Z^{\prime}$ boson to the SM fields correspond to right-handed tops and to left-handed muons as well as neutrinos according to:\footnote{In what follows, for $\eta \sim \mathcal{O}(v)$ we will have $\xi_{T} \sim \mathcal{O}(1)$; consequently, $\varepsilon_{t} \sim \mathcal{O}(v/M_{\mathcal{T}})$.}
\begin{eqnarray}
  \label{eq:Zprime_to_SM}
    g_{Z^{\prime} t_{R}} & = & g_{X} \sin^2 \theta_{R}^{t} = \frac{g_{X}}{ 1+ \xi^2_{\mathcal{T}}} + \mathcal{O}\left( \varepsilon_{t}^2/\xi_{\mathcal{T}}^2 \right) \, , \\
    g_{Z^{\prime} \mu_{L} (\nu)} & = &  g_{X} \sin^2 \theta_{L}^{\mu} = \frac{g_{X}}{1 + \xi^2_{\mathcal{M}}} 
    + \mathcal{O}\left( \varepsilon_{\mu}^2 /\xi_{\mathcal{M}}^2 \right) \, ,
\end{eqnarray}
with $g_{Z^{\prime} t_{L} (\mu_{R})}$ being non-negligible only at order $\varepsilon_{t  (\mu)}^2/\xi_{\mathcal{T(M)}}^2$.
Consequently, integrating out the $Z^{\prime}$ relevantly generates the operator $O^{L u}_{2233}$ with Wilson coefficient:
\begin{equation}
\label{eq:CLu_Zprime}
C^{L u}_{2233} = - \frac{g_{Z^{\prime} t_{R} } g_{Z^{\prime} \mu_{L}} }{m_{Z^{\prime}}^2} \simeq - \frac{1}{(1+ \xi^2_{\mathcal{T}})(1+ \xi^2_{\mathcal{M}}) \, \eta^{2}} \ ,
\end{equation}
together with four-fermion operators built of $t_{R}$ or $\mu_{L},\nu$  fields that can be potentially probed at collider and by experimental signatures like $\nu$-trident production. 

\begin{figure}[tp]
    \includegraphics[scale=0.235]{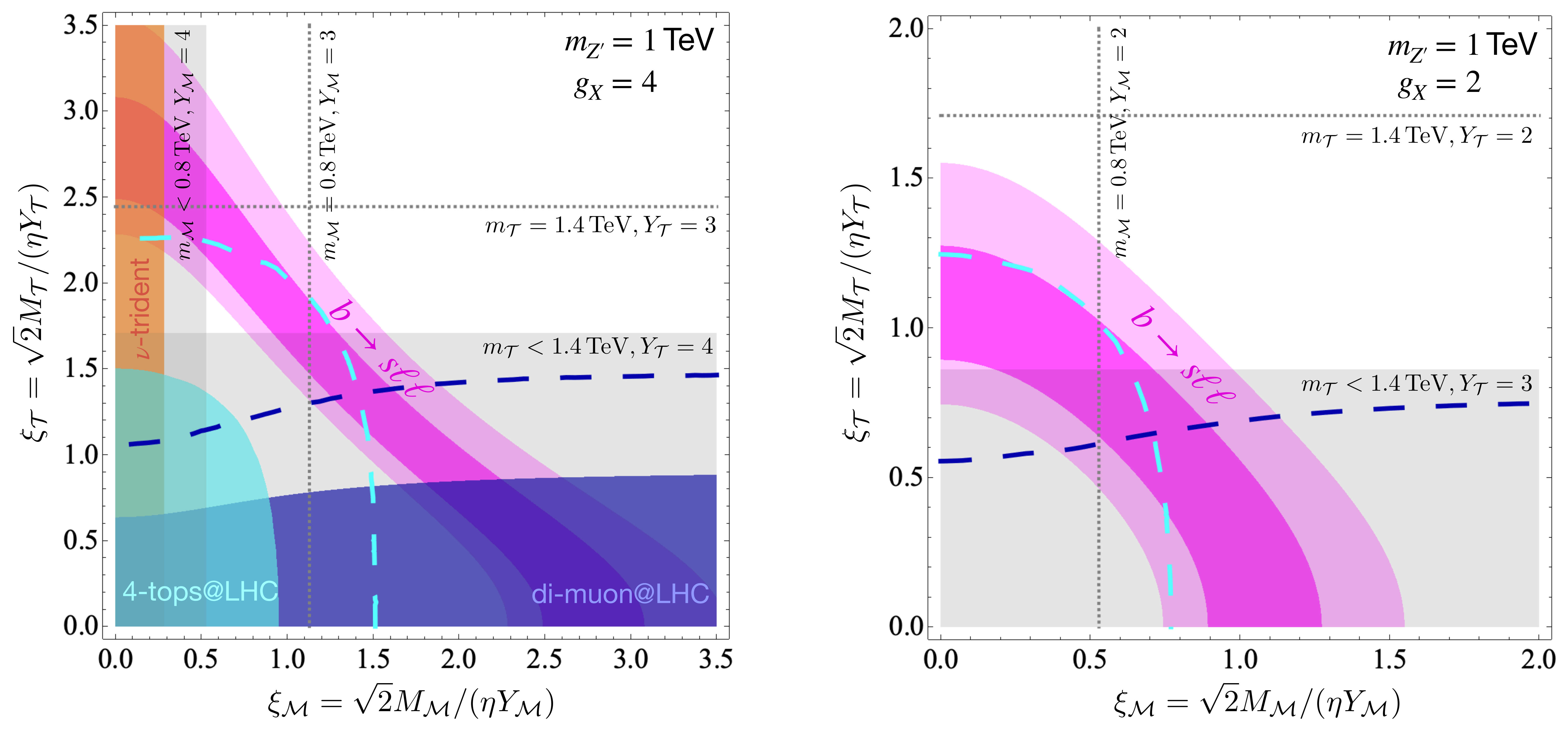}
    \caption{\it 68\% (95\%) probability region in (lighter) magenta for the minimal $Z^{\prime}$ model that addresses $B$ anomalies in the parameter space identified by eq.~\eqref{eq:CLu_Zprime}, with $\eta = m_{Z^{\prime}}/4$ (left panel), and $\eta =  m_{Z^{\prime}}/2$ (right panel), for $m_{Z^{\prime}} = 1~\textrm{TeV}$. Relevant LHC constraints are reported in blue and cyan regions according to the analysis originally performed in ref.~\cite{Camargo-Molina:2018cwu}, together with the corresponding collider projections at 300~fb$^{-1}$. Finally, the gray regions underlie the parameter space where the mass of the vector-like partner lies below current collider limits for a fixed Yukawa coupling as explicitly reported, while dashed lines show the corresponding shift of the limit due to a smaller value of the same type of Yukawa coupling.}    
    \label{fig:Zp_constraints}
\end{figure}

From eq.~\eqref{eq:CLu_Zprime} it is clear that in order to have $|C^{L u}_{2233}| \sim 2 $ TeV$^{-2}$ as highlighted in \autoref{fig:2D_correlations}, one needs to rely on a relatively low symmetry-breaking scale $\eta \lesssim$~TeV;\footnote{Note that even for masses as low as $\mu_{\mathcal{S}} \sim \mathcal{O}(v)$, for $\eta \simeq v$ and $\lambda_{\mathcal{S}} \sim \mathcal{O}(1)$, the interactions of $\mathcal{S}$ do not alter the phenomenology discussed here since the largest $\mathcal{S}$-generated effects are still suppressed as $\mathcal{O}(\varepsilon_{t}^2/\xi^2_{\mathcal{T}})$.} for $m_{Z^{\prime}} \sim$~TeV this implies $g_{X} \gtrsim$~1. In \autoref{fig:Zp_constraints} we show the $1\sigma$ region corresponding to the explanation of $B$ anomalies via eq.~\eqref{eq:CLu_Zprime} in the parameter space $\xi_{\mathcal{T,M}}$, fixing the gauge coupling $g_{X} = m_{Z^{\prime}}/\eta$ for a tentative $Z^{\prime}$ gauge boson at the~TeV scale and the VEV of the new scalar field $\mathcal{S}$ set to $\eta = 250$~GeV and $\eta = 500$~GeV  in the left and right panel, respectively. In the same plot, we re-interpret in our scenario the most relevant collider constraints originally identified in ref.~\cite{Camargo-Molina:2018cwu}. 

For small values of $\xi_{\mathcal{M}}$, the measurement of neutrino-trident production performed in~\cite{Mishra:1991bv} is effective, and its constraint is reported at the 2$\sigma$ level with the orange vertical band.
Under the reasonable assumption that the $Z^{\prime}$ boson is mainly produced at tree level in association with the $ t \bar{t}$ pair, in the blue region we show the 95\% high-$p_{T}$ constraint stemming from the recasting of the $p p \to \mu^{-} \mu^{+} t \bar{t}  $ search at ATLAS~\cite{Aaboud:2017buh}, while in cyan we report the expected constraint on the model from the 4-tops analysis of CMS~\cite{Sirunyan:2017roi}, see ref.~\cite{Camargo-Molina:2018cwu} for further details. From the same work, we also adopt the expected collider constraints for future projected luminosity corresponding to 300~fb$^{-1}$, shown with dashed lines.  Note that these projections become of fundamental importance when it comes to probe the interesting 1$\sigma$ region connected to $B$ anomalies. 
In particular, the right panel in \autoref{fig:Zp_constraints} captures the benchmark for a promising discovery at the High-Luminosity LHC.

Finally, in the same figure, fixing the partner Yukawa coupling to  $\mathcal{O}(1)$ values as reported in the two panels, we mark in gray the region corresponding to the bound on the mass of the vector-like partner expected from collider, taken to be $m_{\mathcal{T}} = 1.4$~TeV from the search at ATLAS in ref.~\cite{Aaboud:2018uek}, and $m_{\mathcal{M}} = 0.8$~TeV from the CMS analysis of ref.~\cite{Sirunyan:2019ofn}.

As already discussed, the scenario depicted in \autoref{fig:Zp_constraints} remains viable under the lens of EW precision as long as we also have some heavy new dynamics yielding at the EW scale an imprint of $O^{HL^{(1)}}_{22}$ consistently with the correlation obtained in the left panel of \autoref{fig:2D_correlations}. 

A simple way to obtain such NP contribution would be to consider the joint effect that the leptonic mixing of the vector-like partner would have together with the kinetic mixing of the $Z^{\prime}$, so far neglected. The $Z$-$Z^{\prime}$ mixing could also originate from charging the new scalar field $\mathcal{S}$ under both Abelian gauge groups, introducing a small misalignment with the standard hypercharge $U(1)_{Y}$ in the UV. However, the required mixing of the $Z^{\prime}$ would end up mediating light-quark pair annihilation into muons: the typical size of the Wilson coefficient of this four-fermion operator would be $\mathcal{O}(g_{Y}^2/m^2_{Z^{\prime}})$, in net tension with the di-muon bound from ATLAS~\cite{Aaboud:2017buh}, probing NP scales as high as $20$~-~$40$~TeV for $\mathcal{O}(1)$ (dimensionless) couplings. Hence, we rule out here this possibility.

Interestingly, it is still possible to generate $O^{HL^{(1)}}_{22}$ without relying on the $Z$-$Z^{\prime}$ mixing, but rather invoking the presence in the UV theory of additional new vector-like leptonic states~\cite{Thomas:1998wy,delAguila:2008pw}. These ones may be phenomenologically interesting in relation  to the problem of the origin of neutrino masses as well as for the prediction of the anomalous magnetic moment $(g-2)_{\mu}$~\cite{Kannike:2011ng}, and may give peculiar multi-lepton signatures at colliders~\cite{Kumar:2015tna,Bhattiprolu:2019vdu}.

In the most economic scenario, we may consider the presence in the UV theory of a pair of new vector-like muonic partners: a singlet of $SU(2)_{L}$, $S_{Y}$, and a triplet of $SU(2)_{L}$, $T_{Y}$, 
where in both cases the subscript $Y$ denotes the hypercharge of the fermion.
These fields would have their own mass terms controlled by the parameters $M_{S_{Y},T_{Y}}$, and interact with the SM doublet $L_{2}$ via the Yukawa couplings $\mathcal{Y}_{S_{Y},T_{Y}}$ according to:
\begin{equation}
    \label{eq:Yuk_singl_tripl}
    \mathcal{Y}_{S_{0}} \bar{S}_{0,R} \tilde{H}^{\dagger} L_{2} + \mathcal{Y}_{T_{0}} \bar{T}^{A}_{0,R} \tau^{A} \tilde{H}^{\dagger} L_{2} + \textrm{h.c.}  \ ,
\end{equation} 
where we have reported the case of vector-like muonic partners with hypercharge $Y=0$. We assume the new Yukawa couplings to be real. Another possibility of interest may be the one of replacing in eq.~\eqref{eq:Yuk_singl_tripl} $\tilde{H} = i \tau^{2} H^*$
with the Higgs doublet, $H$, and involve then the pair of vector-like partners with hypercharge $Y=1$. 

Integrating out these vector-like states from the theory would generate contributions related to $\mathcal{O}^{HL^{(1,3)}}$ ~\cite{delAguila:2008pw,Kannike:2011ng} of the form:
\begin{eqnarray} 
    \label{eq:matching_CHL}
    C^{HL^{(1)}}_{22 } & = &
    \ \ \frac{ \mathcal{Y}^{2}_{S_{0}}}{4 M_{S_{0}}^2} 
    - \frac{ \mathcal{Y}^{2}_{S_{1}}}{4 M_{S_{1}}^2}
    + \frac{ 3 \mathcal{Y}^{2}_{T_{0}}}{4 M_{T_{0}}^2}
    - \frac{ 3 \mathcal{Y}^{2}_{T_{1}}}{4 M_{T_{1}}^2} \ , \\
    C^{HL^{(3)}}_{22} & = & 
    - \frac{ \mathcal{Y}^{2}_{S_{0}}}{4 M_{S_{0}}^2}
    - \frac{ \mathcal{Y}^{2}_{S_{1}}}{4 M_{S_{1}}^2}
    + \frac{ \mathcal{Y}^{2}_{T_{0}}}{4 M_{T_{0}}^2}
    + \frac{ \mathcal{Y}^{2}_{T_{1}}}{4 M_{T_{1}}^2} \ . \nonumber
\end{eqnarray}
Clearly, in order to have $C^{HL^{(1)}}_{22} \sim 0.1$ and negligible $C^{HL^{(3)}}_{22}$\footnote{We have indeed verified that a scenario involving at the same time $C^{Lu}$ and $C^{HL^{(1,3)}}$ would not alter what already highlighted in~\autoref{fig:2D_correlations}, with the best-fit value for $|C^{HL^{(3)}}|$ turning out to be of $\mathcal{O}(10^{-2})$.}, 
one would need to rely on a tuning of the $Y=0$ triplet Wilson coefficient with one of the contributions coming from the singlet vector-like muonic partner. However, once generated at the NP scale $\Lambda \sim \mathcal{O}(M_{T_{0}}) \gg v$, we observe that the relation established between the triplet and singlet contributions to $O^{HL^{(1,3)}}$ would be stable under the RG flow of the SMEFT.

A final comment is needed for the electron scenario reported in the right panel of~\autoref{fig:2D_correlations}, that involves opposite signs for the Wilson coefficients of $O^{Lu}$ and $O^{HL^{(1)}}$ discussed so far. For the former, we note that the sign highlighted in the matching in eq.~\eqref{eq:CLu_Zprime} follows from having assumed the same sign for the charge of the vector-like top and muon partners under $U(1)_{X}$.
 Hence, assuming the vector-like electron partner to have the opposite  $U(1)_{X}$ charge of the top-partner one would be sufficient to  accomplish $C^{Lu}_{1133} > 0$. (Of course, this would also imply a distinct use in eq.~\eqref{eq:new_Yukawa_Mu} of $\mathcal{S}$ and $\mathcal{S}^{\dagger}$ couplings in the Yukawa terms of the vector-like partners involved to keep the theory invariant under $U(1)_{X}$.) For what concerns the generation of $C^{HL^{(1)}}_{11} < 0 $, according to eq.~\eqref{eq:matching_CHL} one needs to correlate once again the contribution stemming from $S_{0}$, or from $S_{1}$, with the effect coming from a $SU(2)_{L}$ triplet, that now needs to be identified with $T_{1}$, namely the triplet of hypercharge $Y=1$.
 
Eventually, we wish also to comment on the possible role of the $O^{eu}$ operator, so far neglected in this discussion, but of potential relevance more in general. In fact, as mentioned earlier, the presence of $O^{eu}$ would be particularly needed in the case where hadronic corrections entering in the amplitude of $B \to K^* \ell \ell $ would be of the size originally estimated in~\cite{Khodjamirian:2010vf}. 
In that case, a solution to flavour anomalies would be preferred in the muonic channel with NP Wilson coefficient $C^{eu}_{2233}$ also substantially deviating from 0, as already discussed in \autoref{sec:GEN_EFT_results}. Then, one would need to involve also the operator $C^{He}_{22}$ to relieve possible tensions with EW precision. In a general picture, the required NP effects from $O^{He}_{11,22}$ can be obtained integrating out heavy vector-like $SU(2)_{L}$ leptonic doublets.

\subsection{Leptoquark scenarios}
\label{sec:mod_Leptoquarks}
An alternative way to reproduce the minimal EFT scenario of \autoref{fig:2D_correlations} would be via \emph{leptoquarks}~(LQ), particles generically predicted in grand unified  theories~(GUTs)~\cite{Pati:1974yy,PhysRevLett.32.438}.
Notoriously, LQ-induced dimension-six operators could be potentially dangerous as they would lead to proton decay at tree level, forcing to push their scale up to the GUT scale. However, the outcome may drastically change in models where the couplings of the LQs would be non-universal with respect to lepton and/or quark flavours. In such a case their mass could be much lower than what typically expected in GUTs and their signatures may actually be probed at present colliders. Interestingly, such LQs are candidates that could explain the lepton flavour universality violation -- even at the loop level here considered~\cite{Camargo-Molina:2018cwu,Coy:2019rfr} -- hinted in the recent LHCb and Belle data. However, this would imply generically a rather non-trivial flavour structure in the theory~\cite{Becirevic:2017jtw}. For a comprehensive survey of LQ models, see for instance~\cite{Buchmuller:1986zs,delAguila:2010mx,Alonso:2015sja,Dorsner:2016wpm,deBlas:2017xtg}. 
 
Here, we limit ourselves to the case of toy models that specifically generate the operators of interest, namely $C^{Lu}_{\ell \ell33}$ and $C^{eu}_{\ell \ell33} $, for $ \ell=1$ (electron)  or $ \ell=2$ (muon). In these peculiar LQ models we then assume that couplings between right-handed top quarks and light leptons are the only ones that actually matter for TeV phenomenology. 

In~\autoref{tab:LQmodels} we list the vector and scalar LQs that constitute the potential LQ candidates able to generate the solutions for $b \to s \ell \ell$ anomalies at one loop under scrutiny.  

\begin{table}[!ht]
	\centering
	{
	\begin{tabular}{ccc }
		\toprule
		 Vector LQ: $\mathcal{V^\mu}$ & $SU(3)_{c} \otimes SU(2)_{L} \otimes U(1)_{Y}$ & Comments \\
		\midrule
       $ \bar L_\ell  \gamma_\mu (\tau^A) Q_3 \, \mathcal{V}^{\mu (A)} $ & $(\overline{\irrep{3}},\irrep{1}\,\text{or}\,\irrep{3}, -2/3 )$ & not of interest \\
        $  ( \mathcal{V^\mu} )^{\dagger} \,\bar e_\ell^c  \gamma_\mu Q_3 $ & $(\overline{\irrep{3}},\irrep{2}\,, 5/6 )$ & not of interest \\
        $\bar{L}^c_\ell \gamma_\mu u_3\, i\tau^2 \, \mathcal{V^\mu}$ & $(\overline{\irrep{3}},\irrep{2}, -1/6 )$ & generates $C^{Lu}_{\ell \ell 33} > 0$\\
         $ \overline {e}_\ell \gamma_\mu u_3\, \mathcal{V^\mu}$ & $(\overline{\irrep{3}},\irrep{1}, -5/3 )$ & generates $C^{eu}_{\ell \ell 33} < 0$ \\    
		\midrule
		 Scalar LQ: $\mathcal{S}$ &  &  \\
		\midrule
      $\bar L_\ell (\tau^A) (i \tau^{2}) \, Q_3^{c} \,\mathcal{S}^{\dagger (A)} $ & $(\overline{\irrep{3}},\irrep{1}\,\text{or}\,\irrep{3} ,1/3 )$ & not of interest  \\    
      $\bar e_\ell Q_3 \,  i \tau^2 \mathcal{S} $ & $(\overline{\irrep{3}},\irrep{2} ,-7/6 )$ & not of interest \\
      $\bar L_\ell u_3\, \mathcal{S} $  & $(\overline{\irrep{3}},\irrep{2} ,-7/6 )$  & generates $C^{Lu}_{\ell \ell 33} < 0$ \\
       $\bar e_\ell^c u_3\, \mathcal{S}$ & $(\overline{\irrep{3}},\irrep{1} ,1/3 )$ & generates $C^{eu}_{\ell \ell 33} > 0$ \\
		\bottomrule
	\end{tabular}
	}
	\caption{Scalar and vector LQ interactions under scrutiny: LQs of interest for our analysis have to generate the dimension-six operators $O^{Lu,eu}_{\ell \ell 33}$.}
	\label{tab:LQmodels}
\end{table}

\begin{figure}[tp]
    \centering 
    \includegraphics[width=0.45\linewidth]{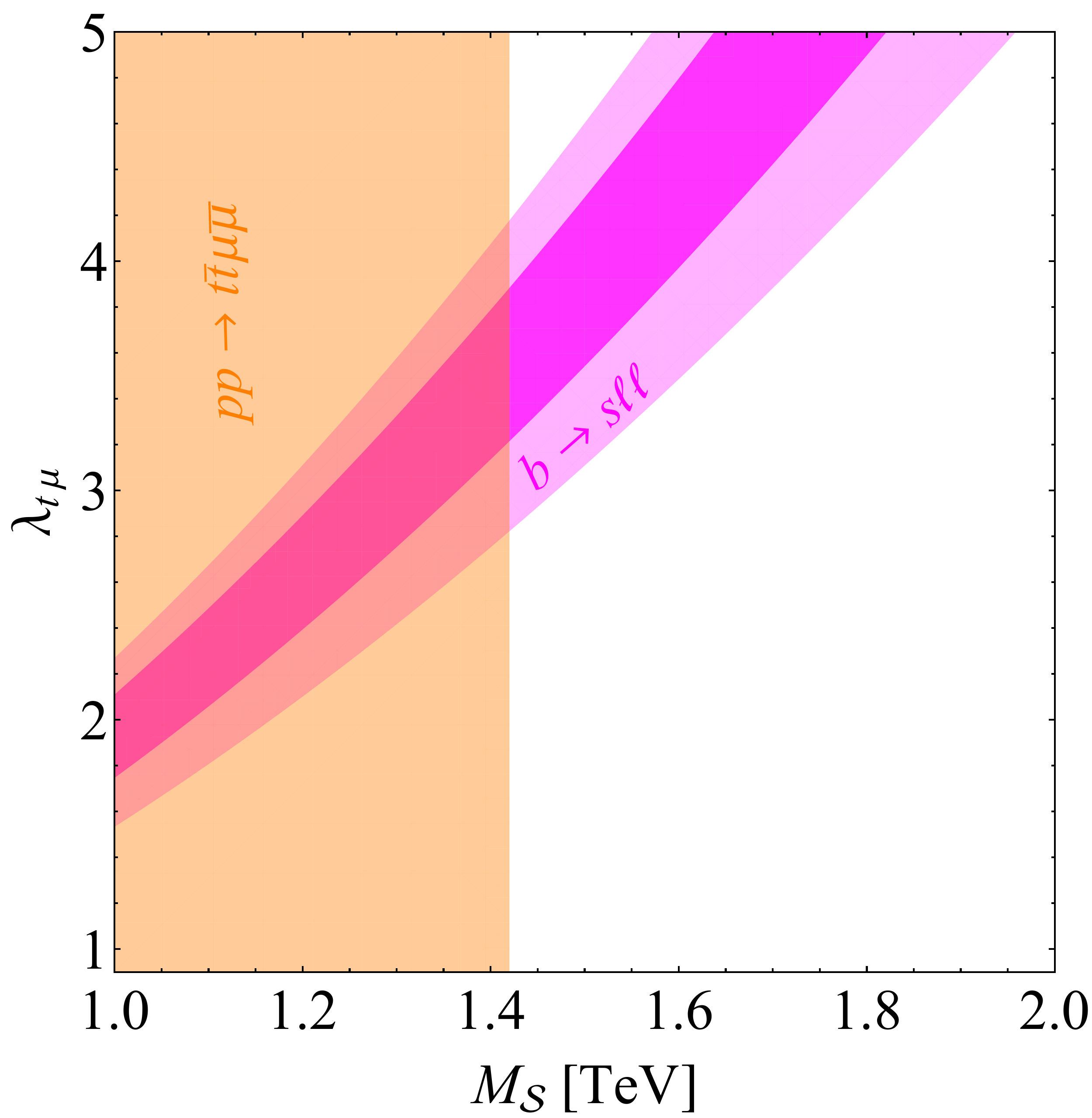}
        \includegraphics[width=0.435\linewidth]{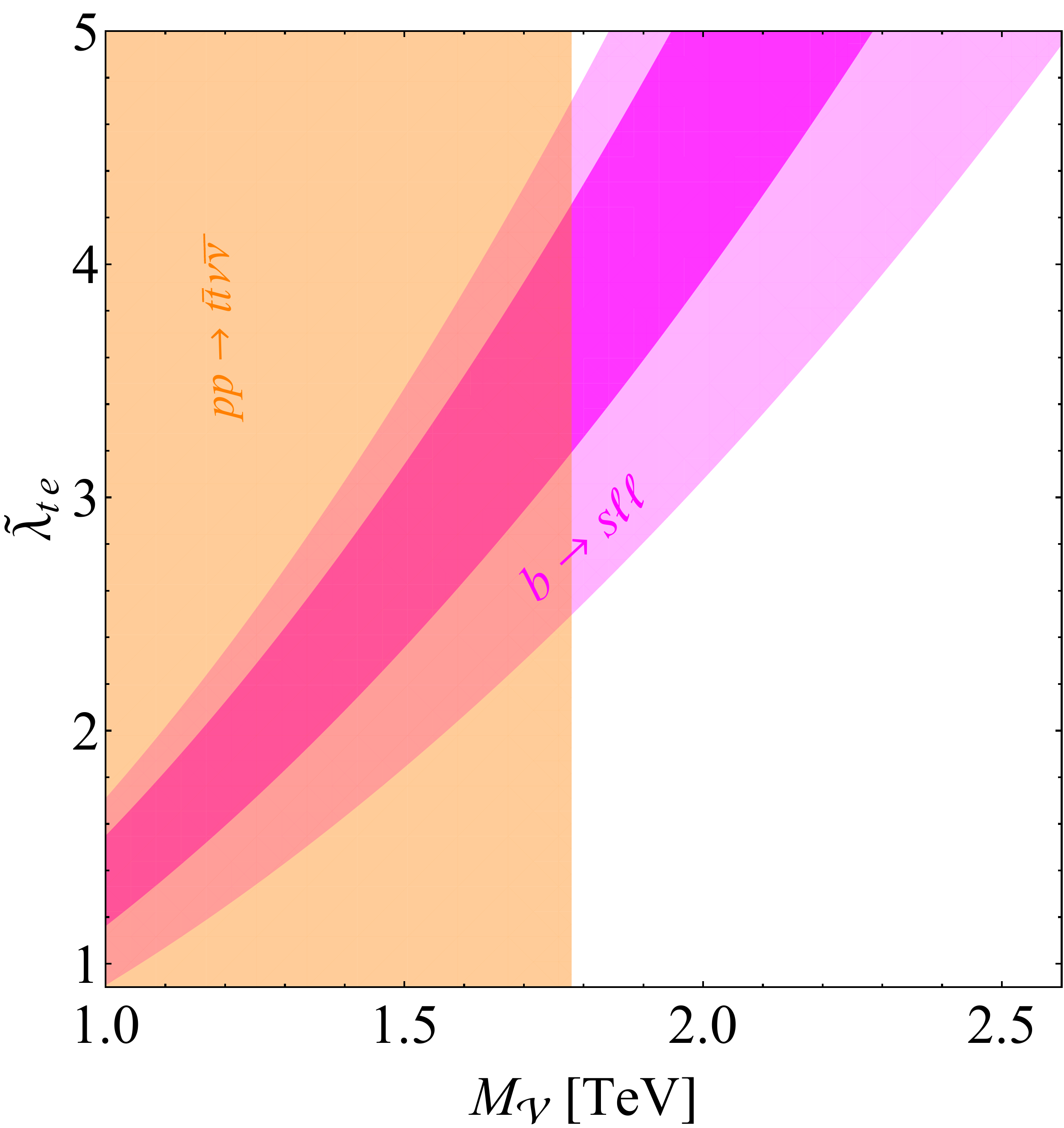}
    \caption{68\% (95\%) probability region in magenta for the LQ candidates addressing $b \to s \ell \ell$ anomalies at one loop. The scalar (vector) LQ corresponds to a solution with LUV effects related to muon (electron) couplings. A conservative bound on the corresponding LQ mass is reported according to the analysis of ref.~\cite{Angelescu:2018tyl}.}    
    \label{fig:LQ_constraints}
\end{figure}

Looking back at \autoref{fig:2D_correlations}, from the table above we recognize as the most economic LQ scenario for the resolution of $B$ anomalies at one loop, the case of the vector LQ $ \mathcal V^\mu\sim(\overline{\irrep{3}},\irrep{2}, -1/6 )$ for LUV effects originating from electron couplings, and the scalar $\mathcal S\sim (\overline{\irrep{3}},\irrep{2} ,-7/6 )$ for the ones associated to muons.
The interaction terms of interest are:
\begin{equation}
\mathcal{L}_{\mathcal V \bar f f} =  \tilde \lambda_{t e}  \, \bar L^c_1\gamma_\mu u_{3} \, i \tau^2 \mathcal V^\mu  + \mathrm{h.c.} 
 \ \ , \ \
\mathcal{L}_{\mathcal S \bar f f} =  \lambda_{t \mu} \,  \bar L_2 u_{3} \mathcal{S}  + \mathrm{h.c.} ,
\end{equation}
leading to the corresponding matching condition:
\begin{equation}
    C^{Lu}_{1133} = +\frac{| \tilde \lambda_{t e}|^2}{M_{\mathcal V} ^2 }  
     \ \ , \ \ C^{Lu}_{2233} = -\frac{|  \lambda_{t \mu}|^2}{2 M_{\mathcal S}^2 } \ .
\end{equation}
In \autoref{fig:LQ_constraints} we report in (lighter) magenta the underlying 1(2)$\sigma$ region where $B$ anomalies are addressed in concordance with the minimal EFT picture of \autoref{fig:2D_correlations}. In the same plot, we also show a conservative estimate of the present LHC constraint on the mass of the LQ states considered, based on the dedicated collider study of ref.~\cite{Angelescu:2018tyl}.

We conclude noting that from the point of view of realizing the economic EFT result in~\autoref{fig:2D_correlations}, these leptoquark models should again be supplied by the combination of a singlet and a triplet $SU(2)_{L}$ muon/electron partners. Otherwise, in these models the leading contribution to $C^{HL^{(1)}}_{\ell\ell}$ would appear only at the loop level, in net distinction with the $Z^{\prime}$ scenario, where the $Z$-$Z^{\prime}$ mixing could be a priori exploited.
  
\section{Summary}
\label{sec:sum}

In this work we have revisited the analysis of $b \to s \ell \ell $ anomalies looking for NP solutions that generate these FCNC processes at one loop and do not involve any new source of flavour violation beyond the SM ones. To this end, we have performed a broad analysis with dimension-six operators in the SMEFT, combining the experimental data on $B$-physics with measurements of EWPO. The general outcome of our study is summarized in~\autoref{fig:ew_flav_bounds} and, supported with \autoref{fig:ew_flav_dg}, shows that a resolution of $B$ anomalies of the MFV nature can be made fully compatible with EW precision.

From the SMEFT results derived we have then proceeded to identifying a minimal EFT scenario as captured in~\autoref{fig:2D_correlations}, that served as a simple guidance for SM UV completions. In this regard, we have explored in some detail the top-phillic and muon/electron-phillic $Z^{\prime}$, interesting for direct searches at collider as highlighted in \autoref{fig:Zp_constraints}. We have also commented on the viable leptoquark scenarios, collected in~\autoref{tab:LQmodels}. For both $Z^{\prime}$ and leptoquark solutions we have found that additional contributions  were necessary in order to maintain $Z$ coupling measurements under control:
in particular, we have shown that a correlated pair of vector-like leptons, a $SU(2)_L$ singlet and a triplet, can realize the minimal EFT scenario depicted on~\autoref{fig:2D_correlations}. We observe that the existence of these particles may be independently motivated by the heavy new dynamics underlying the origin of neutrino masses and/or by a tentative explanation of the $(g-2)_{\mu}$ anomaly~\cite{Kannike:2011ng}.

We conclude by noting that the measurement of $B$ decays at the scale of a few GeV is expected to reach a precision regime with the completion of the future runs at LHC and SuperKEKB. Hence, better measurements of the LUV observables and angular distributions of $b\to s \ell \ell$ will be available in the next few years from Belle~II~\cite{Kou:2018nap} and LHCb~\cite{Bediaga:2018lhg}. These will add a fundamental verification of the current interpretation of $B$ anomalies and of the direction in our search for NP signatures. 
Along these lines, should these signals of LUV persist, their interplay with EW precision measurements could be further tested at future $e^+ e^-$ colliders. In particular, circular $e^+ e^-$ colliders running at the $Z$ pole, such as the FCC-ee~\cite{Abada:2019lih,Abada:2019zxq} or CEPC~\cite{CEPCStudyGroup:2018ghi}, could test deviations in the lepton universality of neutral weak currents with more than one order of magnitude improvement in precision compared to current data. At linear colliders, like the ILC~\cite{Bambade:2019fyw} or CLIC~\cite{deBlas:2018mhx}, where there is no proposed run at the $Z$ pole, it would still be possible to obtain a significant improvement in the measurements of EWPO via radiative return to the $Z$~\cite{Fujii:2019zll}. 
Furthermore, the high-energy regime achievable at linear colliders would allow, after crossing the $t\bar{t}$ threshold, to directly test the effects of the interactions $O^{Lu,eu}_{1133}$ via $e^+ e^- \to t\bar{t}$. 
For the muon case, on the other hand, to test $O^{Lu,eu}_{2233}$ one would still need to rely on more complicated signals, such as $t\bar{t}\mu^+\mu^-$, which would be in any case cleaner than at the LHC. (However, ideal optimal tests of these 4-fermion operators in 2-to-2 scattering processes would require a high-energy muon collider.) All of these could represent valuable additions from a ``flavour'' perspective in the interpretation of EW (and Higgs) measurements at these future machines within the EFT framework~\cite{deBlas:2019rxi,deBlas:2019wgy}.

\acknowledgments
We warmly thank Ramona Gr{\"o}ber, Laura Reina and Luca Silvestrini for valuable feedback. The work of A.A. was in part supported by the MIUR contract 2017L5W2PT. J.B. acknowledges support by the UK Science and Technology Facilities Council (STFC) under grant ST/P001246/1. The work of M.V. is supported by the NSF Grant No. PHY-1915005. This research was supported in part through the Maxwell computational resources operated at DESY, Hamburg, Germany.

\appendix

\section{Discussions on EW fits}
\label{app:EW}

\begin{figure}[t]
    \centering
    \includegraphics[width=\textwidth]{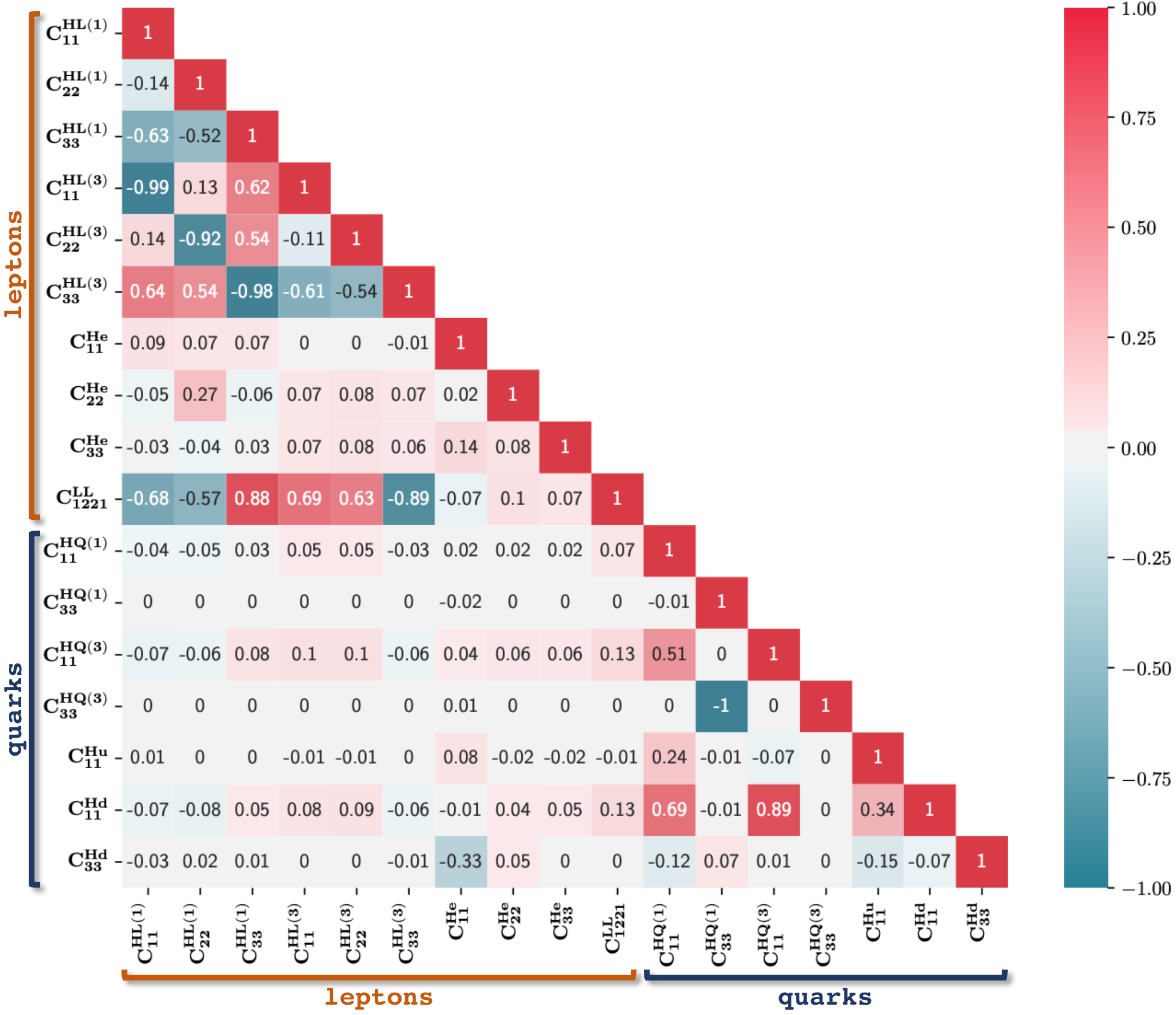}
    \caption{\it The correlation matrix extracted from the SMEFT analysis of the set of independent operators in eqs.~\eqref{eq:SMEFT_op_HL}, \eqref{eq:SMEFT_op_HQ}, \eqref{eq:SMEFT_op_LLLL}, including only their effects at tree-level. The two distinct groups of correlated Wilson coefficients associated to leptonic and quark interactions are remarked as ``leptons'' and ``quarks'', respectively. Note that, compared to \autoref{fig:ew_corr}, in this tree-level analysis there is a significant decorrelation between the constraints on quarks and lepton operators.
    }
    \label{fig:ew_corr_tree}
\end{figure}

Here we revisit the constraints set by EWPO on the parameter space of the SMEFT. We make minimal flavour assumptions and include all quark and lepton operators described in the {\bf EW} fit presented in section~\ref{sec:strategy}. Measurements of EWPO have been extensively studied in the literature~\cite{Han:2004az,delAguila:2011zs,Ciuchini:2013pca,deBlas:2013gla,Falkowski:2014tna,Berthier:2015oma,Efrati:2015eaa,deBlas:2016nqo,deBlas:2017wmn,Ellis:2018gqa,Dawson:2019clf} within the SMEFT framework. The purpose here is to provide further details on the correlation between quark and lepton sectors constrained by EWPO, illustrating some of the effects when going beyond the tree-level analysis. 

The experimental inputs are the same considered for the {\bf EW} fit in section~\ref{sec:strategy}, and include, in particular, the full set measurements taken at LEP/SLD at the $Z$ pole, as well as the measurements of the $W$ boson obtained at LEP II, the Tevatron and the LHC (e.g. mass, width, branching ratios as well as the determination of $\left|V_{tb}\right|$ at the LHC~\footnote{The extraction of $\left|V_{tb}\right|$ could be, a priori, affected by other SMEFT effects entering in single-top production, e.g. 4-fermion operators. Such effects are neglected in our  analysis. The only effect of this input in the EW fits in this paper is to lift a flat direction that would otherwise appear between $C^{HQ^{(1)}}_{33}$ and $C^{HQ^{(3)}}_{33}$, had we excluded this measurement. Even with this input, these two coefficients are nearly $100\%$ correlated, as can be seen in \autoref{fig:ew_corr_tree}.}). For these fits we use the \HEPfit package~\cite{deBlas:2019okz} as for the rest of the work. 

\begin{figure}[t]
    \centering
    \includegraphics[width=\textwidth]{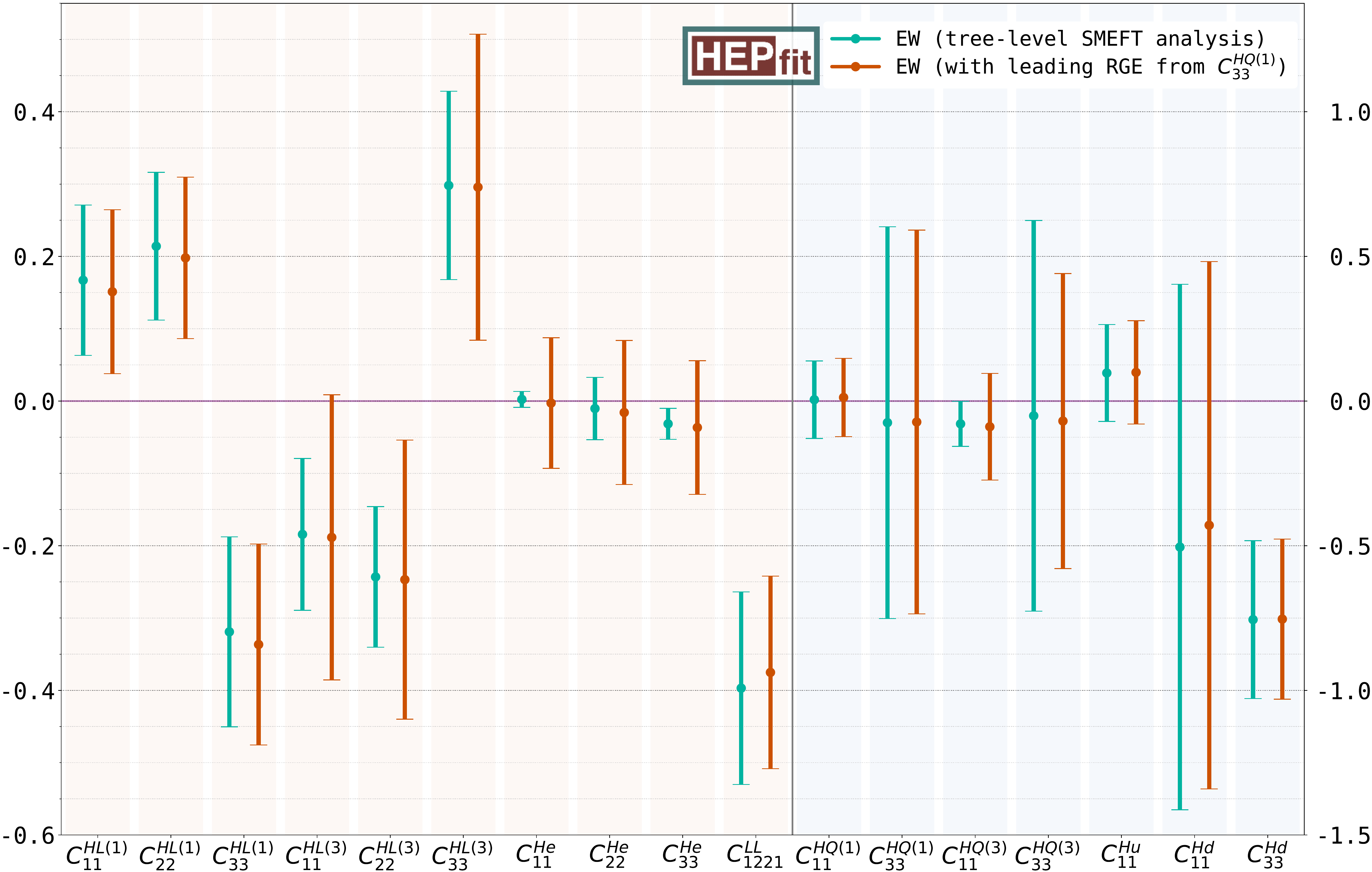}
    \caption{\it Comparison of the mean and standard deviation of the marginalized posterior for the Wilson coefficients (in TeV$^{-2}$) of the operators included in the {\bf EW} fit under two different approximations: in green the results from a pure tree-level analysis; in orange we show the result including the dominant log-enhanced one-loop terms. See text for details. }
    \label{fig:ew_bounds}
\end{figure}

We first consider the case of the {\bf EW} fit at the tree level. In this case, the results of the fit reveal that while there is sizable correlation between the left-handed leptonic operators, as well as between the different quark operators, both sector are however decoupled to a good extent in the fit as can be seen from \autoref{fig:ew_corr_tree}.

For the main fits presented in section~\ref{sec:EFT_results}, however, we also consider the leading logarithmically enhanced contributions at one-loop level via RG running. For our purposes, and considering the size of the bounds on the different operators from the EW fit, the most important contribution comes from $C^{HQ^{(1)}}_{33}$. This induces an universal contribution that propagates into all EWPO. As a result of this, and similar to what was seen between the leptonic operators and the 4-fermion operators due to their interplay in eqs.~\eqref{eq:OLuedRGE}, a non-trivial pattern of correlations between the lepton and quark operator sectors in the {\bf EW} fit arises, as shown in \autoref{fig:ew_corr}. Similar to the change in the bounds on the leptonic operators in the {\bf EW+Flavour} fit once we included the RG effects of the four-fermion operators, the bounds on the leptonic operators also relax in the EW fit once we include the RG effects from $C^{HQ^{(1)}}_{33}$. This is shown in \autoref{fig:ew_bounds}. However, unlike in the {\bf EW+Flavour} fit, such effects do not induce a significant shift in the central values of the Wilson coefficients, which is simply due to the fact that the data selects $C^{HQ^{(1)}}_{33}$ to be centered around zero. 

As can be seen in \autoref{fig:ew_bounds}, the relaxation of the bounds can be in some cases rather dramatic, which brings about the question of what could be the impact of further effects not included in our analysis. We estimated that including the main RG effects for all the other operators in the EW fit amounts to changes of at most $\sim 25\%$. One should also note that finite terms involving the Wilson coefficients of the quark coupling may become relevant at this point. As can be deduced from the full NLO results presented in \cite{Dawson:2019clf}, these are not expected to significantly change the picture. In any case, the overall conclusions on this paper regarding the reconciliation between EW data and $B$ anomalies hold true.

\bibliographystyle{JHEP}
\bibliography{FlavourEW.bib}
\end{document}